%% file: main.tex
\documentclass[sigconf]{acmart}
\usepackage{multirow}
\usepackage{multicol}
\usepackage{booktabs}
\usepackage{subcaption}
\usepackage{caption}
\usepackage{setspace}
\usepackage{amsmath}
\usepackage{tabularx}
\usepackage{balance}
\usepackage{textcomp}
\AtBeginDocument{%
  \providecommand\BibTeX{{%
    \normalfont B\kern-0.5em{\scshape i\kern-0.25em b}\kern-0.8em\TeX}}}
    
\copyrightyear{2021}
\acmYear{2021}
\setcopyright{acmcopyright}
\acmConference[SIGIR '21] {Proceedings of the 44th International ACM SIGIR Conference on Research and Development in Information Retrieval}{July 11--15, 2021}{Virtual Event, Canada.}
\acmBooktitle{Proceedings of the 44th International ACM SIGIR Conference on Research and Development in Information Retrieval (SIGIR '21), July 11--15, 2021, Virtual Event, Canada}
\acmPrice{15.00}
\acmISBN{978-1-4503-8037-9/21/07}
\acmDOI{10.1145/3404835.3462856}
\definecolor{midnightgreen}{rgb}{0.0, 0.29, 0.33}
\definecolor{darkpink}{rgb}{0.91, 0.33, 0.5}

\newcommand{\sa}{$\text{}^{\dagger}$}
\newcommand{\sbx}{$\text{}^{\ddagger}$}
\newcommand{\sab}{$\text{}^{\dagger \ddagger}$}
\newcommand{\sabc}{$\text{}^{\dagger \ddagger \mathsection}$}
\newcommand{\sabcd}{$\text{}^{\dagger \ddagger \mathsection \mathparagraph}$}
\newcommand{\sabd}{$\text{}^{\dagger \ddagger \mathparagraph}$}

\newcommand{\convance}{\texttt{ConvDR}}

\begin{document}
\fancyhead{}

\title{Few-Shot Conversational Dense Retrieval}

\author{Shi Yu$^{1*}$, Zhenghao Liu$^{1*}$, Chenyan Xiong$^2$, Tao Feng$^1$, and Zhiyuan Liu$^1$}\thanks{$*$\ \ indicates equal contribution.}
\affiliation{Tsinghua University$^1$, Microsoft Research$^2$ 
\country{}
} 
\affiliation{
\texttt{\{yus17, liu-zh16\}@mails.tsinghua.edu.cn};
\texttt{chenyan.xiong@microsoft.com};
\country{}
}
\affiliation{
\texttt{fengtao@mail.tsinghua.edu.cn};
\texttt{liuzy@tsinghua.edu.cn}
\country{}
}

\input{abstract}



\maketitle
\input{introduction}
\input{relatedwork}
\input{method}
\input{experiment}

\input{result}
\input{conclusion}

\balance
\bibliographystyle{ACM-Reference-Format}
\bibliography{reference.bib}
\end{document}

%% file: abstract.tex
\begin{abstract}
Dense retrieval (DR) has the potential to resolve the query understanding challenge in conversational search by matching in the learned embedding space. However, this adaptation is challenging due to DR models' extra needs for supervision signals and the long-tail nature of conversational search. In this paper, we present a Conversational Dense Retrieval system, ConvDR, that learns contextualized embeddings for multi-turn conversational queries and retrieves documents solely using embedding dot products. In addition, we grant ConvDR few-shot ability using a teacher-student framework, where we employ an ad hoc dense retriever as the teacher, inherit its document encodings, and learn a student query encoder to mimic the teacher embeddings on oracle reformulated queries. Our experiments on TREC CAsT and OR-QuAC demonstrate ConvDR's effectiveness in both few-shot and fully-supervised settings. It outperforms previous systems that operate in the sparse word space, matches the retrieval accuracy of oracle query reformulations, and is also more efficient thanks to its simplicity. Our analyses reveal that the advantages of ConvDR come from its ability to capture informative context while ignoring the unrelated context in previous conversation rounds. This makes ConvDR more effective as conversations evolve while previous systems may get confused by the increased noise from previous turns. Our code is publicly available at \url{https://github.com/thunlp/ConvDR}.
\end{abstract}

%% file: introduction.tex
\section{Introduction}
Recent advances in language understanding, voice recognition, and availability of edge devices provide search engines an opportunity to better serve users
through \textit{conversational search}~\cite{gao2020recent}.
Instead of ad hoc queries, search in the conversational format enables the system to interact with users through multi-round natural dialogues and take initiatives, to support more complicated information needs and assist with more complex tasks~\cite{aliannejadi2019asking, cast2019overview}.

The multi-round conversational queries yield a unique challenge in conversation search.
Human conversations are contextualized, concise, and assume prior knowledge.
The conversational search usually implies contexts from previous turns with omissions, references, and ambiguities, making it harder for search systems to understand the underlying information needs compared with ad hoc retrieval~\cite{radlinski2017theoretical, trippas2018informing, dalton2020cast}.
Recent conversational search systems address this challenge by reformulating the conversational queries into fully-grown ad hoc queries~\cite{lin2020query, voskarides2020query, vakulenko2020question, yu2020few}.
Learning a reformulated query to mimic the ad hoc query inevitably introduces another noise channel that further aggravates the vocabulary mismatch problem and underperforms their ad hoc counterpart~\cite{yu2020few}.

\input{Figures/fig-model}

A potential solution to fundamentally address the query understanding and vocabulary mismatch challenges in conversational search is to leverage the dense retrieval technique~\cite{lee2019latent, karpukhin2020dense, xiong2020approximate, luan2020sparse, gao2020complementing, khattab2020colbert} recently developed in ad hoc search.
As illustrated in Figure~\ref{fig:intro}, such a system can first encode each conversational query into an embedding that conveys user's information needs, and then directly match with the documents in the learned embedding space~\cite{qu2020orqa}.
Using dense retrieval in conversational search is intriguing. Conversational search aims to model user's information needs from conversation contexts. Thus, it has a more severe vocabulary mismatch problem~\cite{cast2019overview}, which is where dense retrieval thrives on ad hoc search~\cite{xiong2020approximate}.

In practice, however, there is a sharp contradiction between the extreme data-hungriness of dense retrieval~\cite{xiong2020approximate} and the long-tail nature of conversational search~\cite{gao2020recent}.
It is known in ad hoc search that compared to other Neu-IR methods~\cite{mitra2018introduction}, dense retrieval requires more relevance labels, more fine-tuning iterations, and more sophisticated fine-tuning strategies to be useful~\cite{xiong2020approximate, gao2020complementing, luan2020sparse}.
On the other hand, the conversational search may never enjoy the large-scale relevance supervision available in ad hoc search. Its information needs are more complex, more personalized, and often require multiple rounds of queries to develop~\cite{cast2019overview}. Every conversational search session is tail and there are not sufficient training signals to fully supervise current dense retrieval models~\cite{gao2020recent}. 

In this paper, we address this discrepancy with a new few-shot learning approach for conversational dense retrieval (\convance{}). 
We first identify the bottleneck of \convance{} is still at the question representation side, whereas the documents can inherit the same dense representation learned in ad hoc search---their information is still the same.
Then we develop a teacher-student framework to train a student conversational query encoder to ``mimic'' the representation of the oracle query rewrite from the teacher, which is an ad-hoc dense retrieval model~\cite{xiong2020approximate}.
This knowledge distillation (KD) is combined with the standard ranking loss in multi-task learning to enhance the few-shot ability of \convance{}.

In our experiments on TREC CAsT benchmark~\cite{cast2019overview, cast2020overview}, \convance{} shows strong empirical advantages in conversation search. It outperforms the previous state-of-the-art query rewriting based model~\cite{yu2020few} by 9\% and 48\% in retrieval accuracy, achieving even better accuracy than manual question reformulations on CAsT-19. 
In addition, our experiments on QR-QuAC~\cite{qu2020orqa}, a crowd-sourced conversational QA dataset~\cite{choi2020quac, elgohary2019canard}, demonstrate that our knowledge distillation also benefits fully-supervised \convance{} and helps \convance{} almost double the accuracy of previous published state-of-the-arts~\cite{qu2020orqa}.

Our studies find that, compared with question reformulation in the sparse space, the embeddings from our conversational question encoder better approximate the oracle teacher, capture the context in previous turns, and align with the relevant document encodings. Eliminating the extra question reformulation step and directly operating in the dense space not only simplify the pipeline, but also improve online latency as the costly question generation is no longer needed. 

Furthermore, we conduct a series of analyses to understand the effectiveness of \convance{}. We observe that \convance{} effectively and robustly captures the necessary and informative context from previous conversation rounds, while the query rewriter can have dramatically different behavior with alterations in the contexts. 
The ability to handle rich and noisy context information ensures \convance{} to maintain retrieval quality throughout the conversation, while other automatic approaches can get confused in later turns. 

%% file: Figures/fig-model.tex
\begin{figure}[t]
    \centering
    \includegraphics[width=0.95\linewidth]{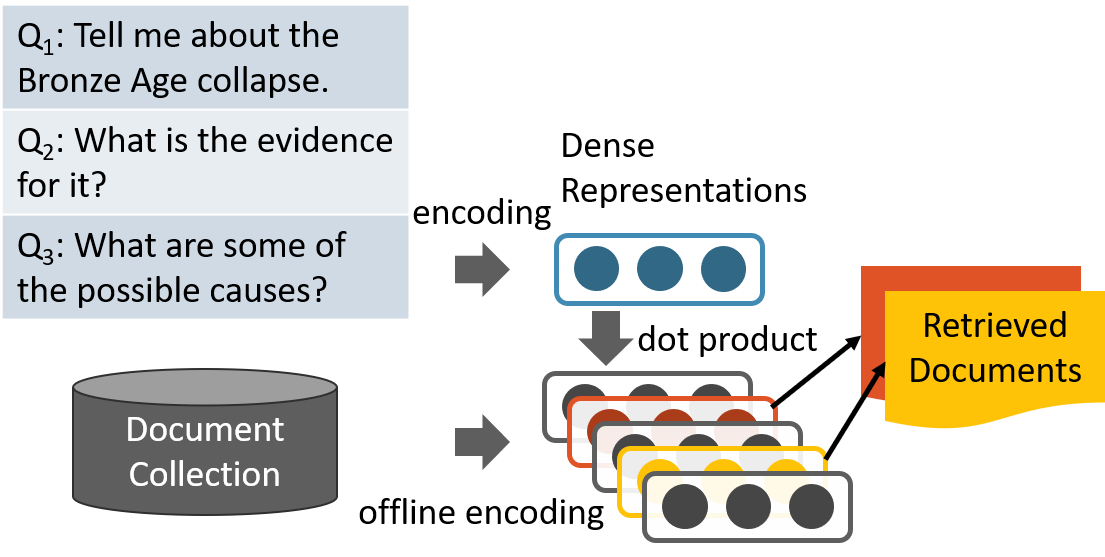}
    \caption{An illustration of Conversational Dense Retrieval. The conversational queries are contextually embedded and matched with documents via dense retrieval. \label{fig:intro}}
    \vspace{-0.3cm}
\end{figure}

%% file: relatedwork.tex
\section{Related Work}

Conversational search is considered as one of the most important research frontiers in information retrieval~\cite{swirl2018}.
With the rapid adoption of smartphones and speakers, 
handling user's search requests in the conversational format is a pressing need in commercial systems~\cite{gao2020recent, croft2019importance}.
Meanwhile, it also provides search engines new opportunities to better support users' complicated information needs with multi-turn interactions and active suggestions~\cite{aliannejadi2019asking, rosset2020leading, zamani2020generating, croft2019importance}.

The TREC Conversational Assistant Track (CAsT) initiative constructed an evaluation benchmark for conversational search~\cite{cast2019overview, cast2020overview}.
Following the voice search scenario in Bing, CAsT formulates a conversational search task where a user's information need is presented by multiple turns of dialogue queries. The task is to retrieve relevant passages for each turn.
The queries are manually curated by the organizers to mimic user behaviors in the voice search scenario (minus the voice recognition part), thus include various human dialogue phenomena: subtle topic shift, reference to the previous context, assuming common knowledge, etc.~\cite{dalton2020cast}.

Recent research on the CAsT benchmark identified the main challenge is on the query side: the document ranking parts can mostly reuse techniques developed in ad hoc search, while the concise conversational queries require new techniques to resolve their context dependency and vocabulary mismatch~\cite{cast2019overview}.
The solutions developed in recent research are to reformulate conversational queries into de-contextualized and fully-grown and ad hoc queries that include all necessary information to represent the user's information needs, and then perform ad hoc retrieval with reformulated queries.

The query reformulation can be conducted via query expansion, i.e., selecting context terms in previous turns to add to the current query's bag-of-word representation, using rules~\cite{lin2020query} or supervised classifiers~\cite{voskarides2020query}.
It can also use query rewriting, which leverages natural language generation models to directly rewrite the conversational queries to de-contextualized ones. 
One of the top-performing systems in TREC CAsT 2019 feeds the previous and current queries to GPT-2~\cite{radford2019language} and uses it to generate fully-grown ad hoc queries. 
Yu et al.~\cite{yu2020few} further fine-tune GPT-2 using synthetic weak supervision data to improve the rewriting accuracy in few-shot scenarios.

These solutions aim to reformulate the conversational query to an ad hoc query in the sparse bag-of-words space and then leverage standard sparse retrieval pipelines such as BM25 and BERT ranker~\cite{cast2019overview, kumar2020making, nogueira2019passage}.
By design, the vocabulary mismatch problem~\cite{croft2010search} in these conversational search systems is more severe than their corresponding ad hoc search systems, as the extra query reformulation step is an additional source of errors. Empirically, there is a notable gap between solutions using automatic query reformulations and those using manual oracle query rewrites~\cite{cast2019overview, yu2020few}.

Recently, dense retrieval has shown strong empirical performance in various ad hoc search and open domain query answering (QA) benchmarks~\cite{bajaj2016ms, nq}.
Instead of sparse word representations, dense retrieval leverages pretrained models, e.g., BERT~\cite{devlin2019bert}, to encode query and documents into embeddings and conduct retrieval purely in the dense representation space~\cite{johnson2019billion}. 
With effective fine-tuning strategies, dense retrieval methods, such as DPR~\cite{karpukhin2020dense}, ME-BERT~\cite{luan2020sparse}, and ANCE~\cite{xiong2020approximate}, significantly outperform sparse retrieval.
On the other hand, learning an effective dense representation space requires more relevance labels and the effectiveness of dense retrieval in few-shot scenarios is less observed~\cite{xiong2020cmt}.

Using the QuAC conversational QA dataset~\cite{choi2020quac, elgohary2019canard}, where crowdsource workers are employed to ask multi-turn questions about a given Wikipedia entity and its description, Qu et al.~\cite{qu2020orqa} constructed a conversational search task OR-QuAC, using the crowded sourced questions as the query, and the evidence passages as the retrieval targets. 
The synthetic nature of OR-QuAC results in all the relevant passages of a dialog reside in the same section of a Wikipedia document, which is not the case in real scenarios. However, OR-QuAC still provides a large number of synthetic labels to showcase the advantage of fully supervised dense retrieval in conversational search~\cite{qu2020orqa}.

%% file: method.tex
\section{Methodology}\label{model}
In this section, we first recap preliminaries in conversational search. Then we present our Conversational Dense Retrieval (\convance{}) framework and its teacher-student few-shot learning strategy.

\subsection{Preliminary}\label{model:preliminary}

The conversational search task~\cite{cast2019overview} is formulated as to find document $d$ from a collection $D$ for each query turn in a multi-round conversation $Q=\{q_i\}_{i=1}^{n}$.
The unique part is that each conversational query $q_k$ can be context-dependent, under-specified, and requires more sophisticated query understanding techniques that infer hidden context from previous queries $Q_{1:k-1}$~\cite{radlinski2017theoretical, trippas2018informing, dalton2020cast}.

A standard solution for conversational search is to reformulate the conversational query into a de-contextualized, fully-grown, ad hoc query~\cite{cast2019overview, yu2020few, vakulenko2020question, lin2020query, lin2020conversational}:
\begin{align}
    q_k; Q_{1:k-1} \xrightarrow{\text{Reformulate}} q'_k, \label{eq:reform}
\end{align}
where the ad hoc query $q'_k$ is expected to fully represent the underlying information needs.
The reformulation can be done by query expansion, for example, selecting terms from $Q_{1:k-1}$ and expanding $q_k$ to $q'_k$~\cite{lin2020query, voskarides2020query}, or rewriting queries, which directly generates the target query $q'_k$ using generation models, e.g., GPT-2~\cite{yu2020few, vakulenko2020question}. 

Recent conversational search benchmarks often provide a manual oracle ad hoc query $q^*_k$ that fully represents the information needs in turn $k$~\cite{dalton2020cast, elgohary2019canard}. The manual oracles are valuable supervision signals to train the query reformulation model in Eq.~\ref{eq:reform}.

After reformulating the conversational query to an ad hoc query, either as a bag-of-words (expansion) or a natural language query (rewriting), the conversational search problem is retrofitted to an ad hoc search problem. 
Then we can reuse the standard retrieval-and-rerank IR pipeline, for example, first use BM25 to retrieval top $K$ documents $D^*$ from the collection $D$:
\begin{align}
     D^* &= \text{BM25} (q'_k, D).
\end{align}
After that, we can use BERT to rerank the candidates $d$ in $D^*$ by computing the ranking score $f^\text{ad}()$ as~\cite{nogueira2019passage}:
\begin{align}
     \mathbf{H}_0(q_k,d) &= \text{BERT} (\text{[CLS]} \circ q'_k \circ \text{[SEP]} \circ  d \circ \text{[SEP]}), \label{eq:rerank:qr} \\
     f^\text{ad}(q'_k,d) &= \text{MLP}(\mathbf{H}_0(q'_k,d)). \label{eq:rerank:mlp}
\end{align}
It uses an MLP layer on the [CLS] embedding ($\mathbf{H}_0$) of BERT that encodes the concatenation ($\circ$) of reformulated $q'_k$ and $d$.

Reformulating conversational search queries to ad-hoc search queries provides a jump start for conversational search~\cite{cast2019overview}, but it also bounds the effectiveness of conversational search systems by ad-hoc search, rather than leveraging the opportunity of multi-turn conversations to provide search experiences of the next generation~\cite{croft2019importance}.
In practice, the query reformulation step is not perfect and further reduces the conversational search accuracy~\cite{yu2020few}.

\input{Figures/fig-method}

\subsection{Conversational Dense Retrieval}\label{model:dense}

Instead of retrofitting to ad hoc retrieval, \convance{} provides an end-to-end solution dedicated to conversational search using dense retrieval~\cite{karpukhin2020dense, xiong2020approximate}.
In general, \convance{} first maps the conversational search queries and documents into one embedding space:
\begin{align}
    \mathbf{q}'_k &\leftarrow g_\text{ConvDR}(q_k; Q_{1:k-1}), \\
    \mathbf{d} &\leftarrow g_\text{ConvDR}(d),
\end{align}
Then \convance{} calculates the retrieval score simply by the dot product of the query embedding and the document embedding:
\begin{align}
    f^\text{DR}(q'_k, d) &= \mathbf{q}'_k \cdot \mathbf{d}.
\end{align}
Top $K$ retrieval in the embedding space is efficiently supported by various libraries~\cite{johnson2019billion}.

As the retrieval function is a simple dot product, the main model capacity is in query and document encoders. 
We use the common BERT-Siamese/Dual-Encoder architecture in dense retrieval~\cite{lee2019latent} to encode the conversational turns and documents:
\begin{align}
     \mathbf{q}'_k &= \mathbf{H}_0(q_k) = \text{BERT}(\text{[CLS]} \circ q_1  \circ \dots \text{[SEP]}  \circ q_k \circ \text{[SEP]}),\label{eq:convdr:q} \\
     \mathbf{d} &= \mathbf{H}_0(d) = \text{BERT}(\text{[CLS]}  \circ d \circ \text{[SEP]}).
\end{align}
We use the BERT [CLS] embeddings as the encodings for the query and document.
Note that we feed in the concatenation of the current query and all previous queries to the query encoder, a prerequisite to capture the information needs in one contextualized embedding.

\convance{} can be optimized to learn retrieval-oriented representations using the negative log likelihood (NLL) \texttt{ranking} loss:
\begin{equation}\label{eq:rank_loss}
    \mathcal{L}_{\text{Rank}}=-\log \frac{\exp \left(\mathbf{q}'_k \cdot \mathbf{d}^+ \right)}{\exp \left(\mathbf{q}'_k \cdot \mathbf{d}^+\right)+\sum\limits_{d^{-} \in D^{-}} \exp \left(\mathbf{q}'_k \cdot \mathbf{d}^-\right)},
\end{equation}
where $d^+$ and $d^-$ are relevant and irrelevant documents for $q_k$.

In the reranking stage, instead of reformulating to query $q'_k$, we directly apply BERT on the concatenation of all queries $Q_{1:k}$ and document $d$: 
\begin{equation}
     \mathbf{H}_0(q_k,d) = \text{BERT} (\text{[CLS]} \circ q_1  \circ \dots \text{[SEP]}  \circ q_k \circ  \text{[SEP]} \circ  d \circ \text{[SEP]}), 
\end{equation}
and then add an MLP layer (Eq.~\ref{eq:rerank:mlp}.) to calculate the ranking score.

\subsection{Few-Shot from Ad Hoc Teacher}\label{model:kd}

A challenge of applying dense retrieval is that, dense retrieval models require more relevance-oriented supervision signals to be effective, even in ad hoc retrieval~\cite{xiong2020approximate, zhan2020learning, karpukhin2020dense}.
In conversational search, \convance{} needs to construct a contextualized query representation using not only the current query but also previous turns, figuring out the context dependency would require additional training signals than encoding an ad hoc query.
Besides, it is harder to accumulate relevance labels in conversational search than in ad hoc search, the information needs in conversational search are expected to be more complicated and personalized~\cite{radlinski2017theoretical}.

To overcome this limitation, we develop a teacher-student framework that improves the few-shot ability of \convance{} by learning from an ad hoc dense retriever.
Given the query embedding $\mathbf{q}_k^*$ obtained from an ad hoc dense retrieval encoder $g_\text{ad hoc}$ on the manual oracle query $q_k^*$ and the document $d$, we use the following teacher-student learning framework to obtain the query and document embeddings in \convance{}:
\begin{align}
    &\mathbf{q}_k^* \leftarrow g_\text{ad hoc}(q_k^*),
    \mathcal{L}_{\text{KD}} = \text{MSE}(\mathbf{q}'_k, \mathbf{q}_k^*),  \label{eq:KD_loss} \\
    &\mathbf{d} \leftarrow g_\text{ad hoc}(d). \label{eq:docreuse}
\end{align}
We use a state-of-the-art ad hoc dense retriever ANCE~\cite{xiong2020approximate} as the teacher. The teacher-student framework introduces two assumptions: 1) the underlying information needs in the manual oracle ${q}^*_k$ and the conversation query $q_k$ are the same thus their embeddings should be the same. 2) the meaningful information in a document $d$ is the same when serving ad hoc search and conversational search, thus its embedding in the two scenarios can be shared.
Following the first assumption, we use MSE loss to distill the knowledge from ANCE to \convance{} (Eq.~\ref{eq:KD_loss})~\cite{hinton2015distilling}. Following the second assumption, we inherit the document embedding from ANCE to \convance{}  (Eq.~\ref{eq:docreuse}).

The knowledge distillation (\texttt{KD}) loss (Eq.~\ref{eq:KD_loss}) is combined with the ranking loss (Eq.~\ref{eq:rank_loss}) in \texttt{multi-task} training:
\begin{align}
    \mathcal{L} = \mathcal{L}_{\text{KD}} + \mathcal{L}_{\text{Rank}}.
\end{align}
To form the negative document set $D^-$ for $\mathcal{L}_{\text{Rank}}$, we retrieve negatives ranked top by ANCE for the manual oracle query $q^*_k$. 

The teacher-student training reduces the need for a large number of relevance labels of \convance{} by inheriting the document representations from ad hoc dense retrieval and mimicking the query embeddings of the manual oracle queries. The ranking loss uses the available conversational search labels to supervise \convance{}, rather than just retrofitting the ad hoc dense retriever. 
The same few-shot paradigms can also be used to train the BERT Reranker.

%% file: Figures/fig-method.tex
\begin{figure}[t]
    \centering
    \includegraphics[width=0.98\linewidth]{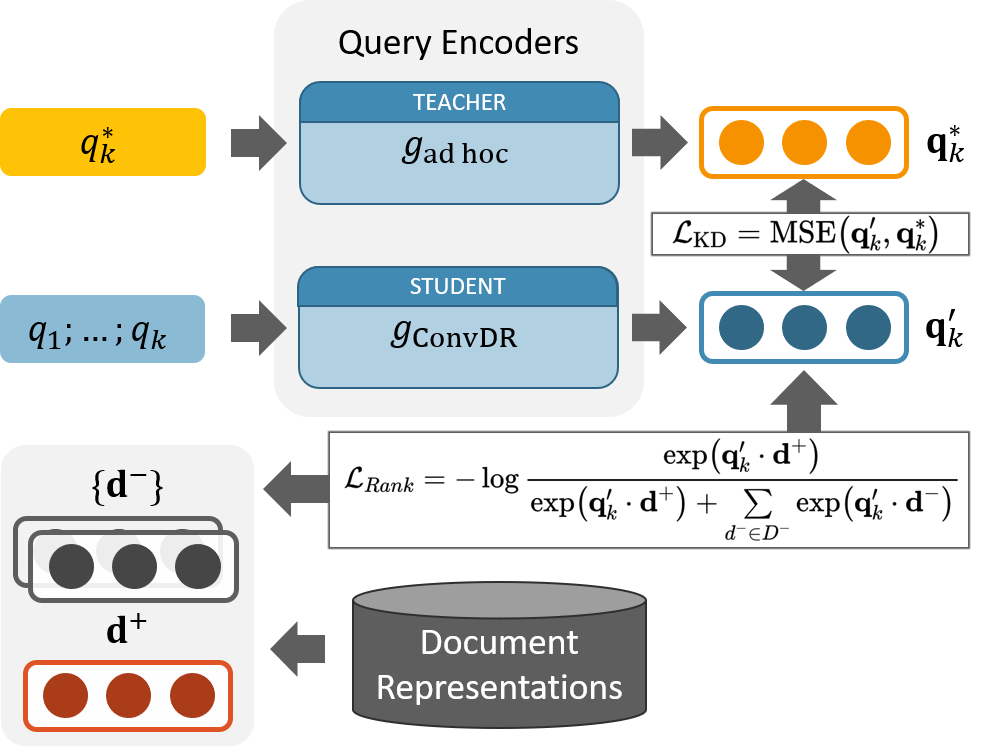}
    \caption{Framework of the \convance{} model. \convance{} encodes the current query $q_k$ and document $d$ as dense representations. Then \convance{} is optimized through the knowledge distillation loss ($\mathcal{L}_{\text{KD}}$) with or without the ranking loss ($\mathcal{L}_{\text{Rank}}$). \label{fig:method}}
\end{figure}

%% file: experiment.tex
\section{Experimental Methodologies}
This section describes our experimental settings, baselines, and implementation details.

\textbf{Datasets.} We use three conversational search benchmarks in our experiments: TREC CAsT 2019~\cite{dalton2020cast} \& 2020~\cite{cast2020overview} include a small amount of TREC-quality relevance labels (few-shot); OR-QuAC~\cite{qu2020orqa} where the search task is synthetic but includes a large amount of labels (``supervised'').
Their statistics are presented in Table~\ref{tab:dataset}.

\input{Tables/tab-dataset}

\textit{CAsT-19.} The TREC Conversational Assistance Track (CAsT) 2019 benchmark~\cite{cast2019overview} provides a series of dialogues (topics) for conversational search. There are 30 training dialogues and 50 test dialogues. Each dialogue contains an average of 9 to 10 queries in natural language forms. Queries in a dialogue are constructed manually to mimic a ``real'' dialogue on a certain topic, with common natural language characteristics including coreferences, abbreviations, and omissions. The later query turns often depend on previous contexts.
The corpus includes passages from MS MARCO~\cite{bajaj2016ms} and TREC Complex Answer Retrieval (CAR)~\cite{dietz2017trec}. 
It provides manual oracle de-contextualized queries for all test queries and relevance judgments for 173 queries in 20 test dialogues. 

\textit{CAsT-20.} CAsT 2020~\cite{cast2020overview} is the dataset of second year of Conversational Assistance Track. It contains 25 evaluation dialogues and uses the same document collection as CAsT-19. The queries can refer to previous answers from system responses, making the dataset more realistic. The organizers included two versions of canonical responses returned by a baseline system: the manual canonical response which uses the manually rewritten query, and the automatic canonical response which uses queries rewritten by a conversation query rewriter~\cite{yu2020few}. 
CAsT-20 provides manually rewritten queries for all queries and relevance judgments for most of them. 

\input{Tables/tab-overall-results-new}

\textit{OR-QuAC}~\cite{qu2020orqa} is a large-scale synthetic conversational retrieval dataset built on a conversational QA dataset, QuAC~\cite{choi2020quac}. 
In QuAC, authors employed crowd-source workers to ask multi-turn questions about a given Wikipedia entity and its description, to synthesize a conversational QA task. CANARD~\cite{elgohary2019canard} recruited crowd-source workers to manually write de-contextualized queries for QuAC.
OR-QuAC extends them to the open-retrieval setting by building a corpus from English Wikipedia (10/20/2019 dump) and using the passages that contain the golden answer as relevant ones.
Note that in OR-QuAC, the answer passages of all turns in a conversation topic are guaranteed to be from the Wikipedia page of the given entity. And the entity is added the first conversation turn~\cite{qu2020orqa}.

\textbf{Evaluation Metrics.} We evaluate model performance on TREC CAsT in terms of MRR\footnote{On CAsT-20, we follow the official evaluation setting~\cite{cast2020overview}, using relevance scale $\ge 2$ as positive for MRR.} and NDCG@3 (main). We also report the average hole rate at cutoff 10 (fraction of the top 10 ranked results without judgments) on CAsT to reveal the uncertainty of the evaluation. For OR-QuAC, we report Recall@5 and MRR@5 following previous work~\cite{qu2020orqa} as well as MAP@10. We conduct the statistical significance test between our implemented methods using permutation test with $p<0.05$.

\textbf{Baselines.} 
We include two groups of baselines, which come from published research and our implementation ones.

The published baselines all use the query reformulation and retrieval-rerank pipeline.
We first include three top performing systems in CAsT-19, \texttt{pgbert}, \texttt{h2oloo\_RUN2}, and \texttt{CFDA\_CLIP\_RUN7}.

\texttt{pgbert}~\cite{cast2019overview} fine-tunes GPT-2~\cite{radford2019language} for query rewriting using the manual rewrites from CANARD~\cite{elgohary2019canard}. It uses Anserini BM25~\cite{yang2017anserini} and BERT reranker. 
\texttt{h2oloo\_RUN2}~\cite{yang2019query} adds topic title in its BM25 retrieval, uses heuristics to expand queries and BERT to rerank. 
\texttt{CFDA\_CLIP\_RUN7}~\cite{yang2019query} adds doc2query~\cite{nogueira2019document} to 
\texttt{h2oloo\_RUN2}.

In addition, we include two recent state-of-the-art systems in CAsT-19. \texttt{RRF (Reranking)}~\cite{lin2020query}\footnote{Ongoing work.} fusions results from \texttt{h2oloo\_RUN2} and Neural Transfer Reformulation (NTR). The latter uses a T5~\cite{t5-model} based query rewriter. 
\texttt{Auto-Rewriter}~\cite{yu2020few} fine-tunes GPT-2 as an automatic query rewriter. Its retrieval and reranking use the standard BM25-BERT pipeline. 

Besides above baselines, on CAsT-20, we include two top performing baselines using the automatic canonical responses:  \texttt{quretecQR} and \texttt{h2oloo\_RUN2$^*$}\footnote{We use an asterisk ($^*$) to distinguish the CAsT-20 run from the CAsT-19 run with the same name.}. \texttt{quretecQR} ~\cite{vakulenko2021leveraging} leverages the QuReTeC~\cite{voskarides2020query} model to classify historical query terms to be added to the current query. \texttt{h2oloo\_RUN2$^*$} ~\cite{cast2020overview} uses heuristics to extract contents from system responses and uses T5 to rewite queries for ranking.
On OR-QuAC, we compare with \texttt{ORConvQA}, the current state-of-the-art~\cite{qu2020orqa}. It is a pipeline neural QA system including a dense retriever, a reranker, and a reader. The dense retriever concatenates the current and historical utterances for retrieval.

In addition to published methods, we implement baselines using different retrieval models, Anserini BM25~\cite{yang2017anserini} or ANCE retrieval~\cite{xiong2020approximate}, and BERT reranker~\cite{nogueira2019passage}.
The queries fed into these models can be raw queries (\texttt{Raw}), automatic query rewrites from a trained GPT-2~\cite{yu2020few} (\texttt{Query Rewriter}), and manual oracle rewrites (\texttt{Manual}, for reference). On CAsT-19, we train GPT-2 following Yu et al.~\cite{yu2020few}, using the Rule-based+CV method that trains the query rewriter on rule-based weak supervision data. On CAsT-20, we train GPT-2 query rewriter on CANARD, with the answer for the previous query. On OR-QuAC, we train the query rewriter on its training set.

\textbf{Implementation Details.} For \convance{} and BERT reranker, we conduct three training paradigms discussed in Section~\ref{model} and include three variants for both: \texttt{KD}, which only distills knowledge from ad hoc models; \texttt{Rank}, which only uses ranking loss; and \texttt{Multi-Task}, the combination of the two.
In addition, we also compare with their zero-shot variants that only trained in ad hoc retrieval.

In the \textit{first stage retrieval}, we use the open-sourced ANCE checkpoints\footnote{https://github.com/microsoft/ANCE} as the teacher~\cite{xiong2020approximate}.
On CAsT, we use its checkpoint on MARCO passages (at the 600k-th step). 
On OR-QuAC, we use its multi-task checkpoint trained on Natural Questions~\cite{nq} and TriviaQA~\cite{triviaqa}; in addition, we continuously train ANCE on the training set of OR-QuAC with manually rewritten queries to adapt it to OR-QuAC. The continuous training uses 3 GPUs, batch size 6, and Adam with a learning rate of 1e-6.

All document embeddings are encoded by ANCE and fixed in our experiments.
The query encoder in \convance{} is warmed up from ANCE and fine-tuned. For the data scarcity on CAsT-20, we further warm up \convance{} on CANARD.
We concatenate historical queries to the current query and discard early turns if exceed 256 tokens. On CAsT-20, we follow the automatic canonical setting and prepend the automatic canonical response to the query of the previous conversational turn. 
All variants of \convance{} are trained with 8 epochs on CAsT-19, 16 epochs on CAsT-20 and 1 epoch on OR-QuAC. For the \texttt{\convance{} (Rank)} and \texttt{\convance{} (Multi-Task)}, we use the manual rewrites to retrieve documents with vanilla ANCE and sample 9 negative documents per query.

\textit{BERT Reranker.} We train BERT rerankers~\cite{nogueira2019passage} on MS MARCO Passage and OR-QuAC tasks with manually reformulated queries. Then we fine-tune BERT rankers with the same three training paradigms, yielding \texttt{BERT (KD)}, \texttt{BERT (Rank)} and \texttt{BERT (Multi-Task)}. We use Reciprocal Rank Fusion (RRF)~\cite{cormack2009reciprocal} to combine the results from the best BERT ranker and \convance{}, yielding \texttt{BERT (RRF)}.

All training on CAsT uses five-fold cross-validation. 
All training processes on OR-QuAC use their official training splits to ensure a fair comparison with previous research~\cite{qu2020orqa}. To train \convance{}, we use 1 GPU, set batch size to 4, and use Adam optimizer with learning rate 1e-5. BERT reranker training uses 4 GPUs, set batch size to 32, and use Adam optimizer with a learning rate of 3e-6.

%% file: Tables/tab-dataset.tex
\begin{table}
  \caption{Data statistics of CAsT and OR-QuAC. The statistics of the judged portion of CAsT are presented in parentheses.}
  \label{tab:dataset}
  \resizebox{0.99\linewidth}{!}{
    \begin{tabular}{l|r|r|rrr} 
    \hline
    \multirow{2}{*}{\textbf{Statistics}}& \textbf{CAsT-19}          & \textbf{CAsT-20}  & \multicolumn{3}{c}{\textbf{OR-QuAC}}  \\
                                & \textbf{Test}                     & \textbf{Test}     & \textbf{Train}& \textbf{Dev}  & \textbf{Test}        \\ 
    \hline
    \# Conversations            & 50 (20)                           & 25 (25)           & 4,383         & 490           & 771         \\
    \# Questions                & 479 (173)                         & 216 (208)         & 31,526        & 3,430         & 5,571       \\
    \# Avg. Question Tokens     & 6.1                               & 6.8               & 6.7           & 6.6           & 6.7         \\
    \# Avg. Questions / Conversation & 9.6                          & 8.6               & 7.2           & 7.0           & 7.2         \\
    \hline
    \# Documents                & \multicolumn{2}{c|}{38M}                              & \multicolumn{3}{c}{11M}  \\
    \hline
    \end{tabular}
   }
\end{table}

%% file: Tables/tab-overall-results-new.tex
\begin{table*}
    \caption{Overall results on TREC CAsT and OR-QuAC. Superscripts $\dagger$, $\ddagger$, $\mathsection$, and $\mathparagraph$ indicate statistically significant improvements over \texttt{BM25-Raw}\sa{}, \texttt{ANCE-Raw}$\text{}^{\ddagger}$, \texttt{BM25-Query Rewriter}$\text{}^{\mathsection}$, and \texttt{ANCE-Query Rewriter}$\text{}^{\mathparagraph}$ or their corresponding reranking results. 
    Unavailable and inapplicable results are marked by  ``n.a.'' and ``--''.
    Best results in each group are marked \textbf{Bold}.
    }
    \small
    \label{tab:result-new}

    \resizebox{0.99\linewidth}{!}{
     \begin{tabular}{l|lll|lll|lll} 
    \hline
    \multirow{2}{*}{\textbf{Method}}        & \multicolumn{3}{c|}{\textbf{CAsT-19}}                 & \multicolumn{3}{c|}{\textbf{CAsT-20}}                 & \multicolumn{3}{c}{\textbf{OR-QuAC}} \\ 
    \cline{2-10}
                                            & \textbf{MRR}  & \textbf{NDCG@3}   & \textbf{Hole@10}  & \textbf{MRR}  & \textbf{NDCG@3}   & \textbf{Hole@10} & \textbf{MAP@10}& \textbf{R@5}  & \textbf{MRR@5}\\ 
    \hline
    \multicolumn{10}{l}{\textbf{Public Baselines}} \\ \hline
     \texttt{pgbert} (from~\cite{cast2019overview}) & 0.665 & 0.413             & n.a.              & --            & --                & --                & --                & --            & --            \\
     \texttt{h2oloo\_RUN2} (from~\cite{cast2019overview}) & 0.714 & 0.434       & n.a.              & --            & --                & --                & --                & --            & --           \\
     \texttt{CFDA\_CLIP\_RUN7} (from~\cite{cast2019overview}) & \textbf{0.715} & 0.436 & n.a.       & --            & --                & --                & --                & --            & --            \\
     \texttt{RRF (Reranking)} (from~\cite{lin2020query}) & n.a. & \textbf{0.536}& n.a.              & --            & --                & --                & --                & --            & --           \\
     \texttt{Auto-Rewriter} (from~\cite{yu2020few}) & n.a.  & 0.492             & n.a.              & --            & --                & --                & --                & --            & --        \\
     \texttt{quretecQR} (from~\cite{vakulenko2021leveraging})& --  & --         & --                & 0.476         & 0.340             & n.a.              & --                & --            & --\\
     \texttt{h2oloo\_RUN2$^*$} (from~\cite{cast2020overview}) & --   & --       & --                & \textbf{0.621}&\textbf{0.493}     & n.a.              & --                & --            & --\\
     \texttt{ORConvQA reranker} (from~\cite{qu2020orqa}) & --   & --            & --                & --            & --                & --                & n.a.              & \textbf{0.314}& \textbf{0.313}\\
    \hline
    \multicolumn{10}{l}{\textbf{First Stage Retrieval Only}} \\ \hline
     \texttt{BM25-Raw}                      & 0.322         & 0.134             & 32.7\%            & 0.160         & 0.101             & 55.9\%            & 0.050             & 0.075         & 0.049 \\
     \texttt{ANCE-Raw}                      & 0.420\sa{}    & 0.247\sa{}        & 60.6\%            & 0.234\sa{}    & 0.150\sa{}        & 68.9\%            & 0.114\sa{}        & 0.149\sa{}    & 0.113\sa{}\\
     \texttt{BM25-Query Rewriter}           & 0.581\sab{}   & 0.277\sa{}        & 19.0\%            & 0.250\sa{}    & 0.159\sa{}        & 39.7\%            & 0.200\sab{}       & 0.302\sab{}   & 0.202\sab{}\\
     \texttt{ANCE-Query Rewriter}           & 0.665\sabc{}  & 0.409\sabc{}      & 43.1\%            & 0.375\sabc{}  & 0.255\sabc{}      & 49.6\%            & 0.451\sabc{}      & 0.584\sabc{}  & 0.457\sabc{}\\
     \convance{}                            & \textbf{0.740}\sabcd{} & \textbf{0.466}\sabcd{} & 40.3\%& \textbf{0.501}\sabcd{}& \textbf{0.340}\sabcd{}& 42.2\%& \textbf{0.607}\sabcd{} & \textbf{0.750}\sabcd{} & \textbf{0.616}\sabcd{} \\
    \hline
    \multicolumn{10}{l}{\textbf{With BERT Reranker}} \\ \hline
     \texttt{BM25-Raw\textrightarrow BERT-Raw}          & 0.503\sbx{}   & 0.306\sbx{}       & 36.1\%            & 0.220         & 0.154             & 48.7\%            & 0.154             & 0.171         & 0.157       \\
     \texttt{ANCE-Raw\textrightarrow BERT-Raw}          & 0.449         & 0.275             & 50.7\%            & 0.249         & 0.184\sa{}        & 56.9\%            & 0.174\sa{}        & 0.207\sa{}    & 0.177\sa{} \\
     \texttt{BM25-QR\textrightarrow BERT-Query Rewriter}& 0.780\sab{}   & 0.495\sab{}       & 21.1\%            & 0.381\sab{}   & 0.264\sab{}       & 30.9\%            & 0.537\sab{}       & 0.615\sab{}   & 0.554\sab{} \\
     \texttt{ANCE-QR\textrightarrow BERT-Query Rewriter}& 0.747\sab{}   & 0.481\sab{}       & 28.4\%            & 0.383\sab{}   & 0.298\sabc{}      & 36.7\%            & 0.604\sabc{}      & 0.710\sabc{}  & 0.615\sabc{} \\
     \texttt{\convance{}\textrightarrow BERT}           & \textbf{0.802}\sabd{} & 0.526\sabd{} & 33.0\%         & 0.506\sabcd{} & 0.362\sabcd{}     & 32.1\%            & \textbf{0.758}\sabcd{} & \textbf{0.850}\sabcd{} & \textbf{0.773}\sabcd{} \\
     \texttt{\convance{}\textrightarrow BERT (RRF)}     & 0.799\sabd{}  & \textbf{0.541}\sabcd{}& 31.2\%        & \textbf{0.545}\sabcd{}& \textbf{0.392}\sabcd{}& 31.1\%& 0.698\sabcd{}     & 0.821\sabcd{} & 0.718\sabcd{} \\
    \hline
    \multicolumn{10}{l}{\textbf{Manual for Reference}} \\
    \hline
     \texttt{BM25-Manual}                   & 0.671         & 0.347             & 10.5\%            & 0.445         & 0.301             & 17.7\%            & 0.316             & 0.430         & 0.330      \\
     \texttt{ANCE-Manual}                   & 0.740         & 0.461             & 37.1\%            & 0.591         & 0.422             & 32.2\%            & 0.507             & 0.642         & 0.515  \\
     \texttt{BM25-Manual\textrightarrow BERT-Manual}    & 0.842         & 0.548             & 16.0\%            & 0.632         & 0.470             & 7.1\%             & 0.645             & 0.719         & 0.665     \\
     \texttt{ANCE-Manual\textrightarrow BERT-Manual}    & \textbf{0.835}& \textbf{0.566}    & 23.1\%            & \textbf{0.663}& \textbf{0.483}    & 14.9\%            & \textbf{0.683}    & \textbf{0.778}&\textbf{0.698}\\
    \hline
    \end{tabular}}

\end{table*}

%% file: result.tex
\section{Evaluation Results}
\label{result}
In this section, we present five groups of experiments on the overall performance of \convance{}, its few-shot strategies, the learned query representations, its ability to model contexts, and case studies.

\input{Tables/tab-efficiency}

\subsection{Overall Performance}\label{result:overall}

The results on CAsT and OR-QuAC are listed in Table~\ref{tab:result-new}.

\textbf{Few-Shot Accuracy.} On CAsT, \texttt{\convance{}} outperforms all baselines in \textit{first stage retrieval}. It improves \texttt{BM25-Query Rewriter} by 68\% and 113\%, showing the advantage of dense retriever over BM25 sparse retrieval.
\convance{} is more effective than \texttt{ANCE-Query Rewriter} and also simpler: It directly maps queries to the embedding space, rather than intermediately reformulating queries in the sparse space. Noteworthy, \convance{} shows high hole rates on CAsT, indicating that the performance may be underestimated. 

Encouragingly, \convance{} performs on par with \texttt{ANCE-Manual} on CAsT-19, and the gap between automatic and manual conversational search systems on CAsT-20 is also greatly eliminated. In later experiments, we study how \convance{} mimics the manual oracle query in the representation space (Sec.~\ref{sec:qrepexp}) while also effectively capturing salient context information (Sec.~\ref{sec:contextexp}).

Solely using embedding dot products, \convance{} already outperforms the best participating system in CAsT-19, \texttt{CFDA\_CLIP\_RUN7}. The latter is a well-designed system with state-of-the-arts sparse retrieval and Neu-IR techniques~\cite{yang2019query}. We have made promising progress since the CAsT initiative.

The effectiveness of \convance{} in first stage retrieval can translate to the reranking stage with BERT reranker. On CAsT-19, \texttt{\convance{}\textrightarrow BERT} outperforms the query rewriting baselines. Combined with the results from \convance{} using RRF, it further outperforms the previous state-of-the-art, \texttt{RRF (Reranking)}~\cite{lin2020query}, which uses additional query rewriting supervision from CANARD~\cite{elgohary2019canard} and a more powerful pretrained model, T5. On CAsT-20, \texttt{\convance{}\textrightarrow BERT} outperforms every baseline by a large margin except \texttt{h2oloo\_RUN2}, which uses a dense-sparse hybrid retrieval model followed by a powerful T5 ranking model with T5-based query reformulation. We also observe a larger automatic-manual gap in reranking compared to the first stage retrieval when reranking dense retrieval candidates. 
Reranking on the top of dense retrieval require more future research.

\textbf{Supervised Accuracy.} On OR-QuAC where a large amount of synthetic relevance labels are available, our fully-supervised \convance{} outperforms all previous methods by huge margins. With first stage retrieval only, \convance{} almost doubles the accuracy of \texttt{ORConvQA reranker}, the best system from Qu et al.~\cite{qu2020orqa}; \texttt{ConvDR\textrightarrow BERT} even outperforms \texttt{ANCE-Manual\textrightarrow BERT-Manual}. 

Empirical gains on OR-QuAC should be read with caution because the synthetic data includes various artifacts that over-simplify the retrieval problem~\cite{qu2020orqa}\footnote{Such artifacts exist in many benchmarks. They do not disqualify the value of these benchmarks in progressing research as long as we are aware of their limitations.}. For example, all relevant passages for a dialog are from the Wikipedia page of the entity in the first round. Nevertheless, the synthetic data helps us take a peek at the potential of \convance{} when we have large-scale user feedback signals when search engines become conversational.
\input{Tables/tab-ablation}
\input{Figures/fig-rep-q}
\input{Figures/fig-rep-qd}

\textbf{Efficiency.} We measure the online inference efficiency of \convance{} and compare with the spare solutions in Table~\ref{tab:efficiency}.

In production systems, one can aggressively batch queries (Batched) in dense retrieval and enjoy the speed up with tensor operations~\cite{xiong2020approximate}. Nevertheless, we also measure the efficiency when queries are processed individually for a more complete picture. In per query setting, BM25 and ANCE have similar latency, but query rewriting~\cite{yu2020few} is significantly more expensive than query encoding. 
The rewriting uses costly language generation where tokens are sampled one-by-one, while encoding uses one forward pass per query.

\input{Figures/fig-round}
\input{Figures/fig-overlap}

\subsection{Effectiveness of Training Strategies}
\label{sec:ablation}
In this experiment, we study the performance of different training strategies in \convance{} and BERT Reranker on CAsT-19 and OR-QuAC in Table~\ref{tab:ablation}. The models include \texttt{Zero-Shot}, which uses no specific training on conversational search; \texttt{KD}, which only uses the knowledge distillation (MSE Loss, Eq.~\ref{eq:KD_loss}); \texttt{Rank}, which only uses the ranking loss (Eq.~\ref{eq:rank_loss}); and \texttt{Multi-Task} which uses both.

\textbf{Effectiveness in Few-Shot Scenarios.} On CAsT 2019, \texttt{KD} is the only one that effectively trains \convance{}. The other three perform worse than plain ANCE results on CAsT 2019. The relevance supervision, either used individually in \texttt{Rank} or combined in \texttt{Multi-Task}, downgrades the model to worse performance than \texttt{Zero-Shot}.
Current dense retrievers require a large amount of relevance supervision signals or may perform worse than BM25. Our teacher-student few-shot learning is necessary for the effectiveness of \convance{}. 
In comparison, the BERT reranker seems to be more robust and can benefit from the labels from TREC CAsT. Combined with the KD loss, \texttt{BERT (Multi-Task)} can achieve the best accuracy.  

Similar trends are also observed on the per query comparisons with \texttt{Manual} runs. Notably, in the retrieval stage, \texttt{KD} is the closest to \texttt{ANCE-Manual} with most ties. \convance{} mimics its teacher well. 

\textbf{Effectiveness in Supervised Setting.} With a large number of training signals from OR-QuAC, training directly using \texttt{Rank} works well with both \convance{} and BERT reranker. Still our teacher-student learning with \texttt{KD} provides addition signals when combined in \texttt{Multi-Task} and helps \convance{} become the best on OR-QuAC.

\subsection{Learned Query Representations}
\label{sec:qrepexp}
In this experiment, we analyze the learned query and document embeddings from \convance{} and ANCE variants.

\textbf{Query Embeddings.} 
We first plot the average similarity (dot product) between the embeddings of the same query learned by different methods in Figure~\ref{fig:query-sim}. 
As expected, \texttt{ANCE-Raw} and \texttt{ANCE-Manual} are very different. 
\texttt{0S} is different from all the others including \texttt{Raw}. Directly applying the ad hoc dense retriever on the concatenation of multi-turn queries does not lead to meaningful embeddings.
\texttt{ANCE-Query Rewriter} is similar to both \texttt{ANCE-Manual} \texttt{(MQ)} and \texttt{ANCE-Raw}. The automatic query rewrites are in between the raw queries and the manual oracle ones.
Without explicitly reformulating a new query, \texttt{\convance{} (KD)} effectively approximates \texttt{ANCE-Manual} by end-to-end learning in the embedding space.

We further demonstrate this effect by using t-SNE~\cite{van2008visualizing} to plot a conversation topic in Figure~\ref{fig:query-tsne}. The \texttt{Zero-Shot} embedding resides in the middle of the raw embeddings of the current (Q6) and previous turns; 
treating the concatenation of all queries as one ad hoc query yielded average embeddings.
 \texttt{\convance{} (KD)} is as close to the current query Q6 as its teacher \texttt{MQ}, but balances its location with previous queries. 
 Rather than the manual oracle query,
\texttt{\convance{} (KD)} has access to the raw queries and can capture information of previous turns. 
 We provide an example of this effect in Sec.~\ref{sec:case}.

\textbf{Similarities with Document Embeddings.} 
The effects on the query embeddings also influence their similarities with relevant documents. As show in Figure~\ref{fig:qd-sim}, \convance{} \texttt{(KD)} has the highest similarity, even higher than \texttt{ANCE-Manual} \texttt{(MQ)}. The automatic query rewriting, however, has lower similarity compared with \texttt{MQ}.
Mimicking the manual oracle queries in the sparse word space is a harder task than in the dense space.

We also visualize the learned query embedding of the 6th turn of topic 79 with their document embeddings in Figure~\ref{fig:qd-tsne}. The zero-shot query embedding resides in the middle of irrelevant documents. 
The other three all locate closer to relevant documents in the space. Among them, \texttt{ANCE-Query} \texttt{Rewriter} seems to introduce some biases and favors a subset of relevant documents. The rewritten query might only capture a part of the full information needs. 
\texttt{ANCE-Manual} and \texttt{\convance{} (KD)}, in comparison, position the query closer to the middle of the relevant documents. In representation learning, such a more uniformed representation allocation often indicates better generalization ability~\cite{wang2020understanding}.

\input{Tables/tab-case-new}
\subsection{Effectiveness in Capturing Contexts}
\label{sec:contextexp}
In this experiment, we study the effectiveness of \convance{} on learning query representations from multi-turn conversations.

\textbf{Accuracy Through Multiple Turns.} We first study the behavior of dense retrieval models at different conversation turns, using their per turn retrieval accuracy and their query embedding similarities between conversation rounds.

As shown in Figure~\ref{fig:per-turn}, the accuracy of \texttt{ANCE} variants drops as the conversation goes on, the same with previous conversational search systems~\cite{yu2020few, cast2019overview}. The queries in later conversation turns are more likely to depend on previous turns, while the increased reasoning space---conversation turns---makes the context dependency harder to resolve.
However, \texttt{\convance{} (KD)} maintains its accuracy throughout the conversation and performs on par with the Manual Oracle (\texttt{MQ}) on all turns.
This further confirms that our few-shot strategy helps \convance{} capture the context information required to understand user's information needs.

\convance{}'s better query understanding ability is also reflected in Figure~\ref{fig:turn-sim}. The query representations from \convance{} \texttt{(KD)} are more similar with each other in adjacent rounds, indicating a smooth transition during conversation turns in the embedding space.
In comparison, the de-contextualized manual queries (\texttt{MQ}) are less correlated, as \texttt{ANCE-Manual} does not use previous turns. 
The embeddings from automatic rewritten queries change more dramatically during the conversation.
Adding query rewriting as an intermediate step introduces new source of error. 
Learning the contextualized query embeddings directly is more effective and efficient.

\textbf{Dependency on Previous Conversation Turns.}
To further study how \convance{} models contexts in previous turns, we conduct an intrusion test by randomly discarding a previous turn in the conversation, and track the changes of query embeddings before and after. 
We compare this embedding change with the term overlap between the discard turn and the manual oracle query of the current turn. The term overlap is an approximation of the context importance in the discarded turn. The results are plotted in Figure~\ref{fig:overlap}.

\convance{} \texttt{(Zero-Shot)} is the least impacted by this intrusion. It does not capture much context information anyway and removing a previous turn does not make it worse.
\texttt{ANCE-Query} \texttt{Rewriter} is significantly impacted by this intrusion test.
Language generation models are known to be fragile and a small change of its input can lead to dramatically different outputs~\cite{yu2020few, radford2019language, Holtzman2020The}.

\texttt{\convance{} (KD)} shows a smooth correlation between the embedding changes and the term overlap of the context turn and the manual oracle. It learns to capture the important context (indicated by high term overlap with oracle) and attend less on unrelated context.
Its embedding changes are also more subtle: a lot of query embeddings are changed but most only by a little, showing its robustness to context removal. A context term may appear in multiple previous turns and an effective model should be able to use duplicated contexts to recover from removing one turn if possible.
\texttt{\convance{} (Multi-Task)} has more variances in this intrusion test. The ranking loss with limited relevant labels is not reliable.

\subsection{Case Study}
\label{sec:case}

Our last study shows some winning examples of \convance{} in Table~\ref{tab:case}.

In the first case, the raw query is not explicitly context-dependent and can be an ad hoc query itself. The manual oracle is the same with the raw query. However, the acronym ``ACL'' is ambiguous and by encoding with previous conversational turns, \convance{} disambiguates it correctly.
The second case is a typical coreference error made by automatic query rewriters. \texttt{\convance{}} correctly identifies the right context in its latent space and retrieves the correct passage.

The last case is interesting as it reflects the ``conservative'' behavior of the automatic query rewriter. When it is not able to resolve the context dependency, the rewriter tends to withdraw to and repeat the raw query~\cite{yu2020few}. 
By modeling all previous conversational turns, \texttt{\convance{}} provides a more meaningful query representation by attending more to the first turn, as indicated by the score drops, and capturing the salient context terms ``Darwin'' in its retrieval.

%% file: Tables/tab-efficiency.tex
\begin{table}
  \caption{Efficiency of \convance{} and baselines in our university-standard computing environment. ANCE retrieval is measured on GPUs by FAISS~\cite{johnson2019billion}.}
  \label{tab:efficiency}

    \begin{tabular}{l|r|r} 
        \hline
        \textbf{Operation}      & \textbf{CAsT-19}      & \textbf{OR-QuAC}      \\
        \hline
        Query Rewriting      & 245ms                 & 400ms                 \\
        Query Encoding       & 15ms                  & 15ms                 \\
        \hline
        BM25 Retrieval          & 645ms                 & 266ms                 \\
        ANCE Retrieval          & 545ms                 & 312ms                 \\
        ANCE Retrieval (Batched)& 2.4ms               & 1.1ms                 \\
        \hline          
        \texttt{BM25-Query Rewriter} Total&890ms                 & 666ms                 \\
        \texttt{ANCE-Query Rewriter} Total&805ms                 & 727ms                 \\
        \convance{} Total       & 560ms                 & 327ms                 \\
        \convance{} Total (Batched)&17.4ms            & 16.1ms                \\
        \hline
    \end{tabular}
\end{table}

%% file: Tables/tab-ablation.tex
\begin{table}
    \caption{Results of different training paradigms. Candidates from \texttt{ConvDR (KD)} and \texttt{ConvDR (Multi-Task)} are reranked with BERT on CAsT-19 and OR-QuAC. Win/Tie/Loss (\%) are compared with \texttt{ANCE-Manual} or \texttt{ANCE-Manual\textrightarrow BERT-Manual}.
    }
    \label{tab:ablation} 
\resizebox{0.99\linewidth}{!}{
    \begin{tabular}{l|cc|cc} 
    \hline
    \multirow{2}{*}{\textbf{Method}}    & \multicolumn{2}{c|}{\textbf{CAsT 2019}}& \multicolumn{2}{c}{\textbf{OR-QuAC}}   \\ \cline{2-5}
                                        & \textbf{NDCG@3}   & \textbf{W/T/L}    & \textbf{MRR@5}    & \textbf{W/T/L}      \\ 
    \hline
    \multicolumn{5}{l}{\textbf{First Stage Retrieval Only}}                                                          \\ 
    \hline
    \texttt{\convance{} (Zero-Shot)}    & 0.202             & 13/30/57          & 0.568             & 24/61/16\\
    \texttt{\convance{} (KD)}           & \textbf{0.466}    & 38/39/24          & 0.519             & 19/63/18\\
    \texttt{\convance{} (Rank)}         & 0.084             & 3/19/78           & 0.588             & 29/52/19\\
    \texttt{\convance{} (Multi-Task)}   & 0.157             & 12/19/69          & \textbf{0.616}    & 30/56/14\\
    \hline
    \textbf{With Rerank}                & \multicolumn{2}{c|}{\convance{} (KD)}                 &  \multicolumn{2}{c}{\convance{} (Multi-Task)}  \\ 
    \hline
    \texttt{BERT (Zero-Shot)}           & 0.407             & 16/22/62          & 0.760             & 19/68/12\\
    \texttt{BERT (KD)}                  & 0.497             & 39/17/44          & 0.769             & 20/68/12\\
    \texttt{BERT (Rank)}                & 0.513             & 36/16/48          & \textbf{0.773}    & 21/68/11\\
    \texttt{BERT (Multi-Task)}          & \textbf{0.526}    & 38/18/45          & 0.772             & 20/68/11\\ 
    \hline
    \end{tabular}}
\end{table}

%% file: Figures/fig-rep-q.tex
\begin{figure}[t]
    \centering
        \begin{subfigure}[t]{0.49\columnwidth}
        \centering
        \includegraphics[width=\linewidth]{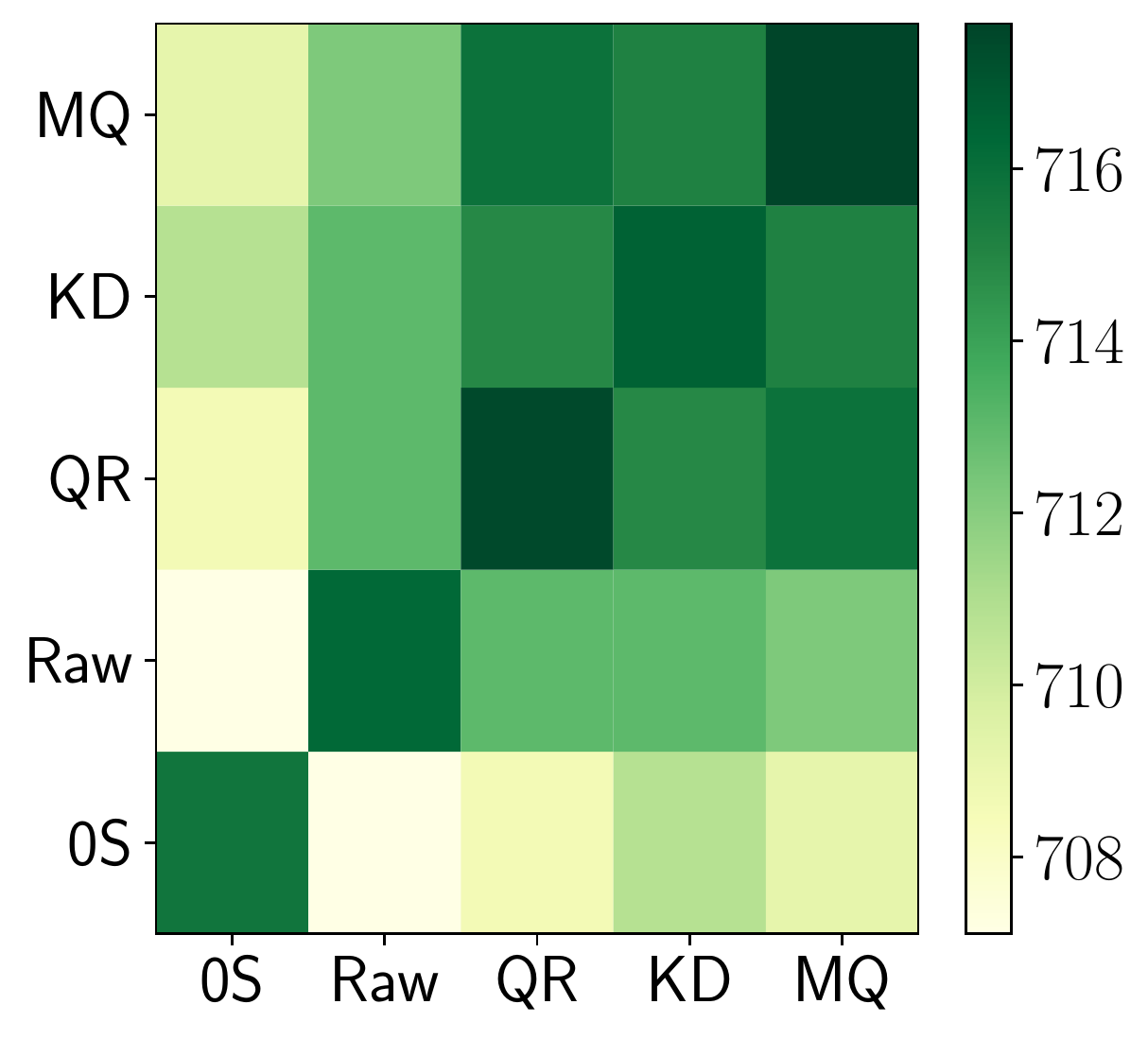}
        
        \caption{Similarity Scores.\label{fig:query-sim}}
    \end{subfigure}
        \begin{subfigure}[t]{0.45\columnwidth}
        \centering
        \includegraphics[width=\linewidth]{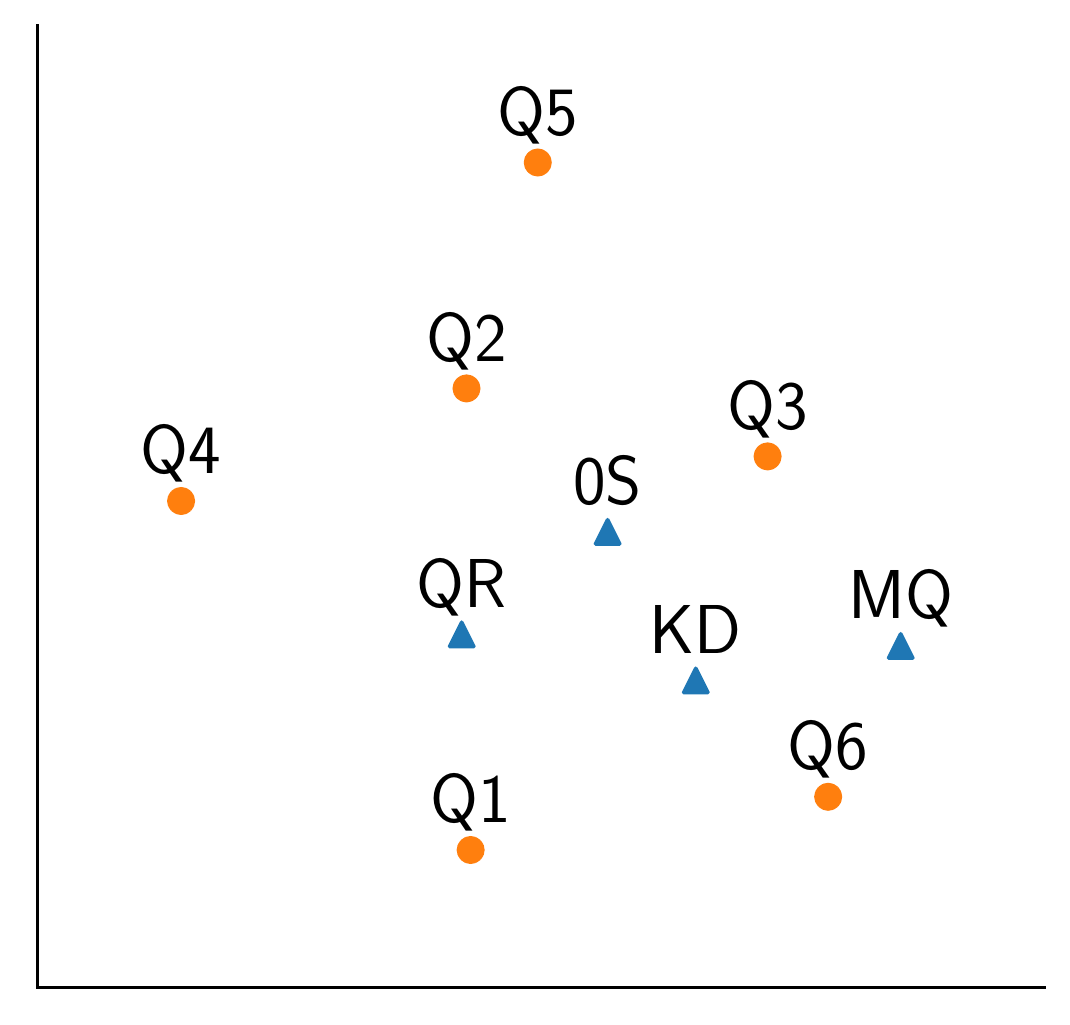}
        \caption{Query Representations.\label{fig:query-tsne}}
    \end{subfigure}

    \caption{Query representations on CAsT-19.
     Query embeddings are from \texttt{\convance{} (Zero-Shot)} \texttt{(0S)}, \texttt{ANCE-Raw} \texttt{(Raw)}, \texttt{ANCE-Query} \texttt{Rewriter} \texttt{(QR)}, \convance{} \texttt{(KD)} and \texttt{ANCE-Manual} \texttt{(MQ)}.
      The similarity (a) is the average dot product score of two representations of the same query.
    The Figure (b) plots the query embeddings at the 6-th turn of topic 79. Q1-Q6 are query embeddings from \texttt{ANCE (Raw)}.}
    \label{fig:query-reps}
\end{figure}

%% file: Figures/fig-rep-qd.tex
\begin{figure}[t]
    \centering
    \vspace{-0.3cm}
    \begin{subfigure}[t]{0.49\columnwidth}
        \centering
        \includegraphics[width=\linewidth]{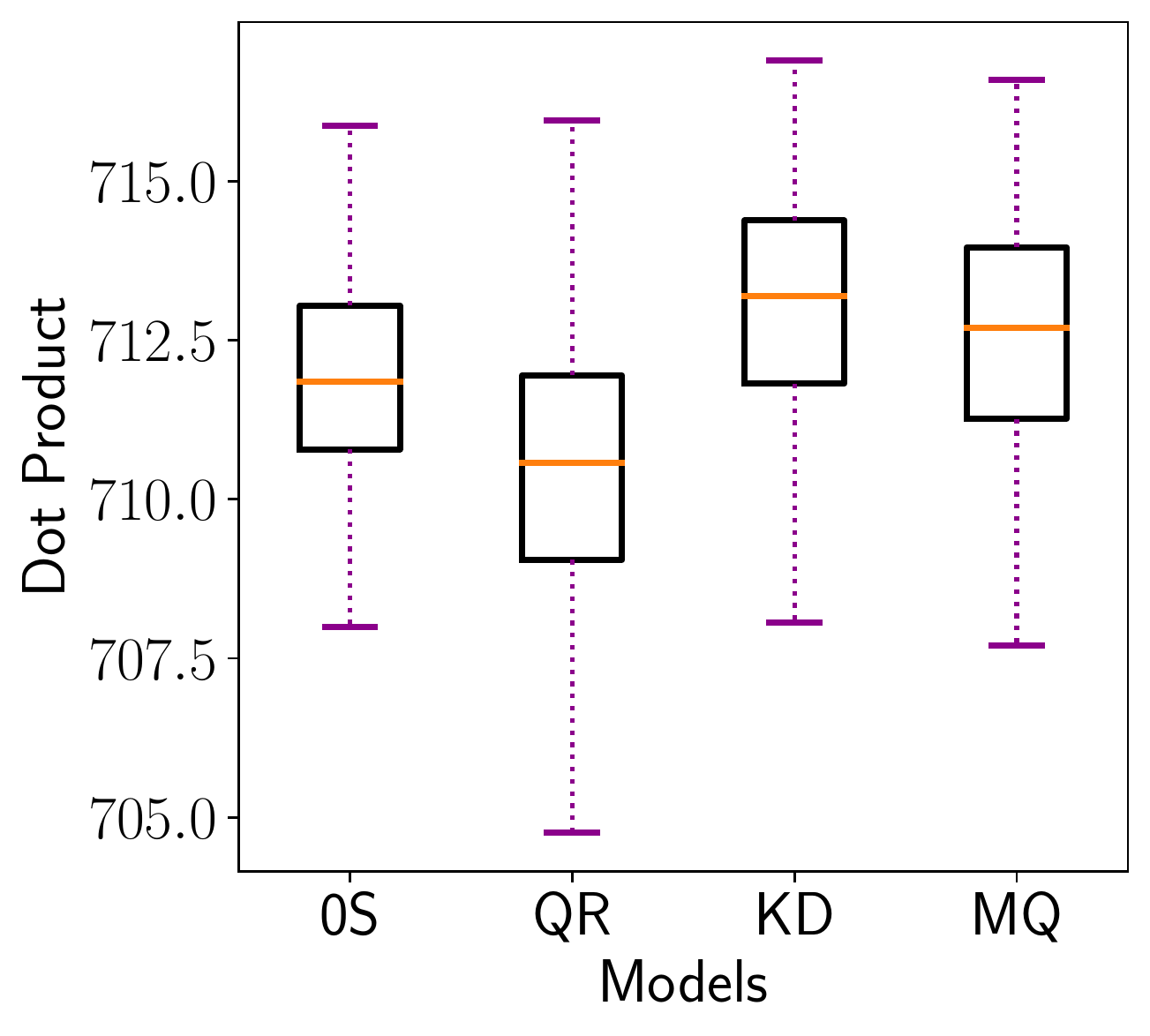}
       
        \caption{Q-D similarity. \label{fig:qd-sim}}
    \end{subfigure}
    \begin{subfigure}[t]{0.45\columnwidth}
        \centering
        \includegraphics[width=\linewidth]{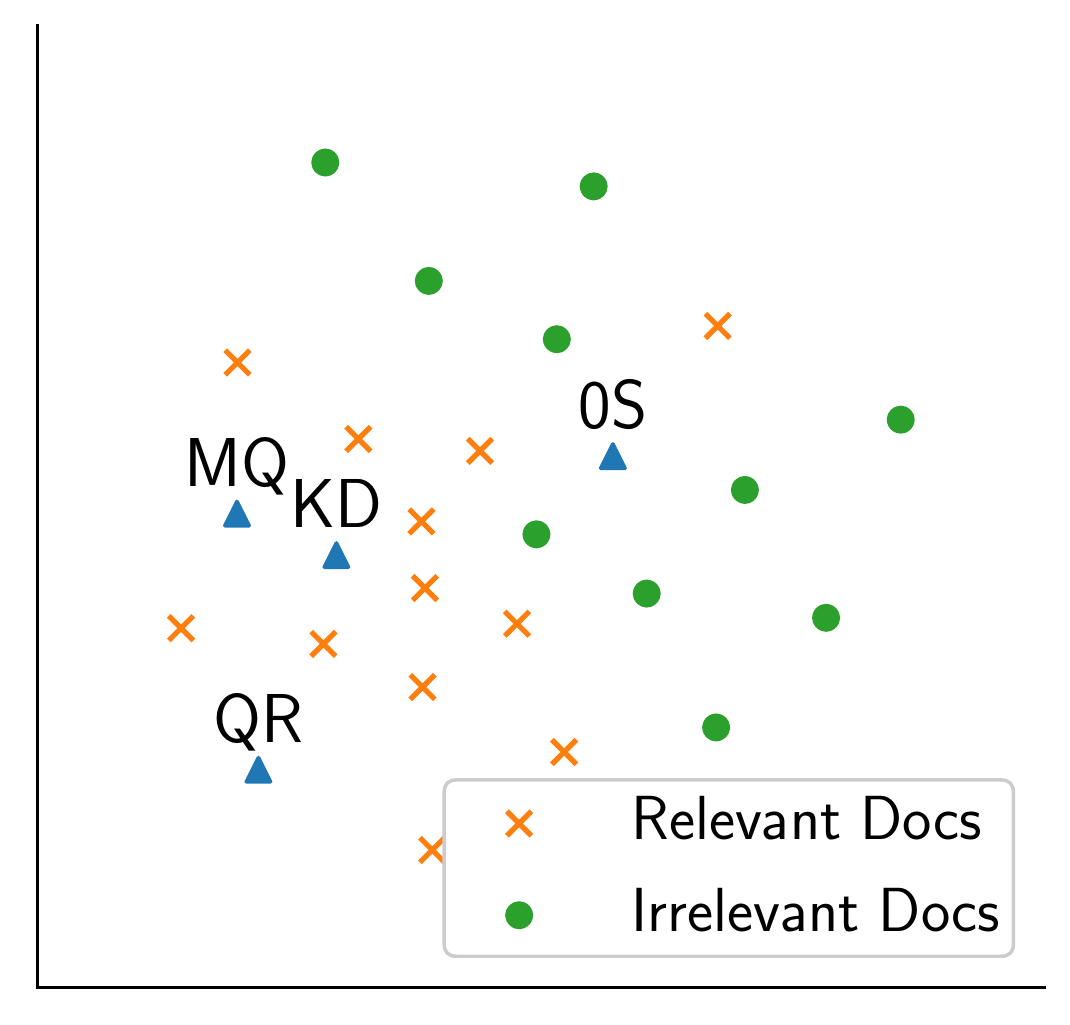}
        \caption{T-SNE representations. \label{fig:qd-tsne}}
    \end{subfigure}
    \caption{
    Similarities of query and document embeddings on CAsT-19. Query embeddings are from \texttt{\convance{} (Zero-Shot)} \texttt{(0S)}, \texttt{ANCE-Query} \texttt{Rewriter} \texttt{(QR)}, \texttt{ANCE-Manual} \texttt{(MQ)} and \texttt{\convance{} (KD)}. The similarity (a) is the average dot product score of queries and their nearest relevant documents. The representation distribution of queries and documents from turn 6, topic 79 of CAsT-19 is plotted through t-SNE in (b).
    \label{fig:qd}}
    \vspace{-0.3cm}
\end{figure}

%% file: Figures/fig-round.tex
\begin{figure}[t]
    \centering
    \begin{subfigure}[t]{0.485\columnwidth}
        \centering
        \includegraphics[width=\linewidth]{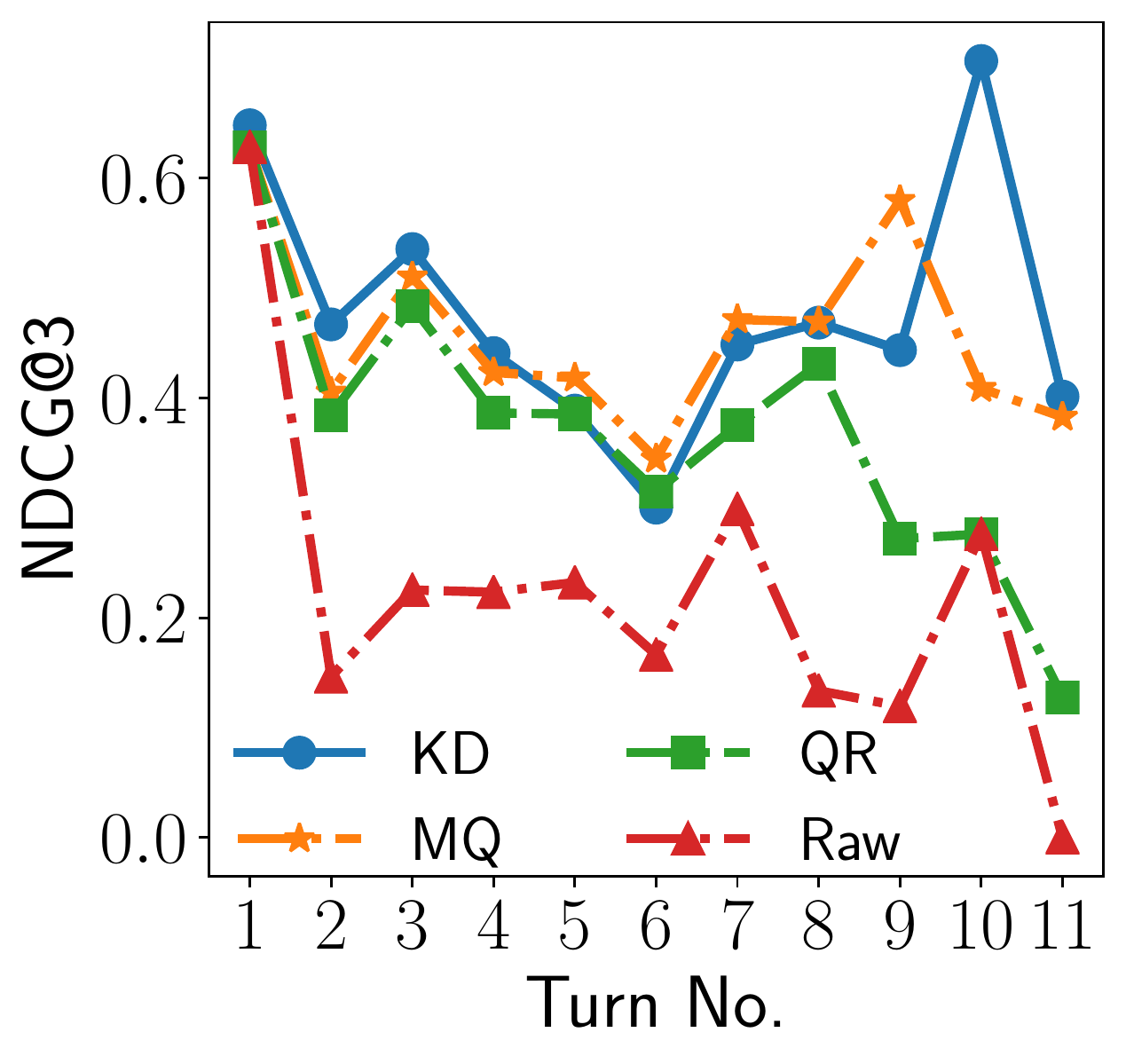}
       
        \caption{Ranking accuracy. \label{fig:per-turn}}
    \end{subfigure}
    \begin{subfigure}[t]{0.49\columnwidth}
        \centering
        \includegraphics[width=\linewidth]{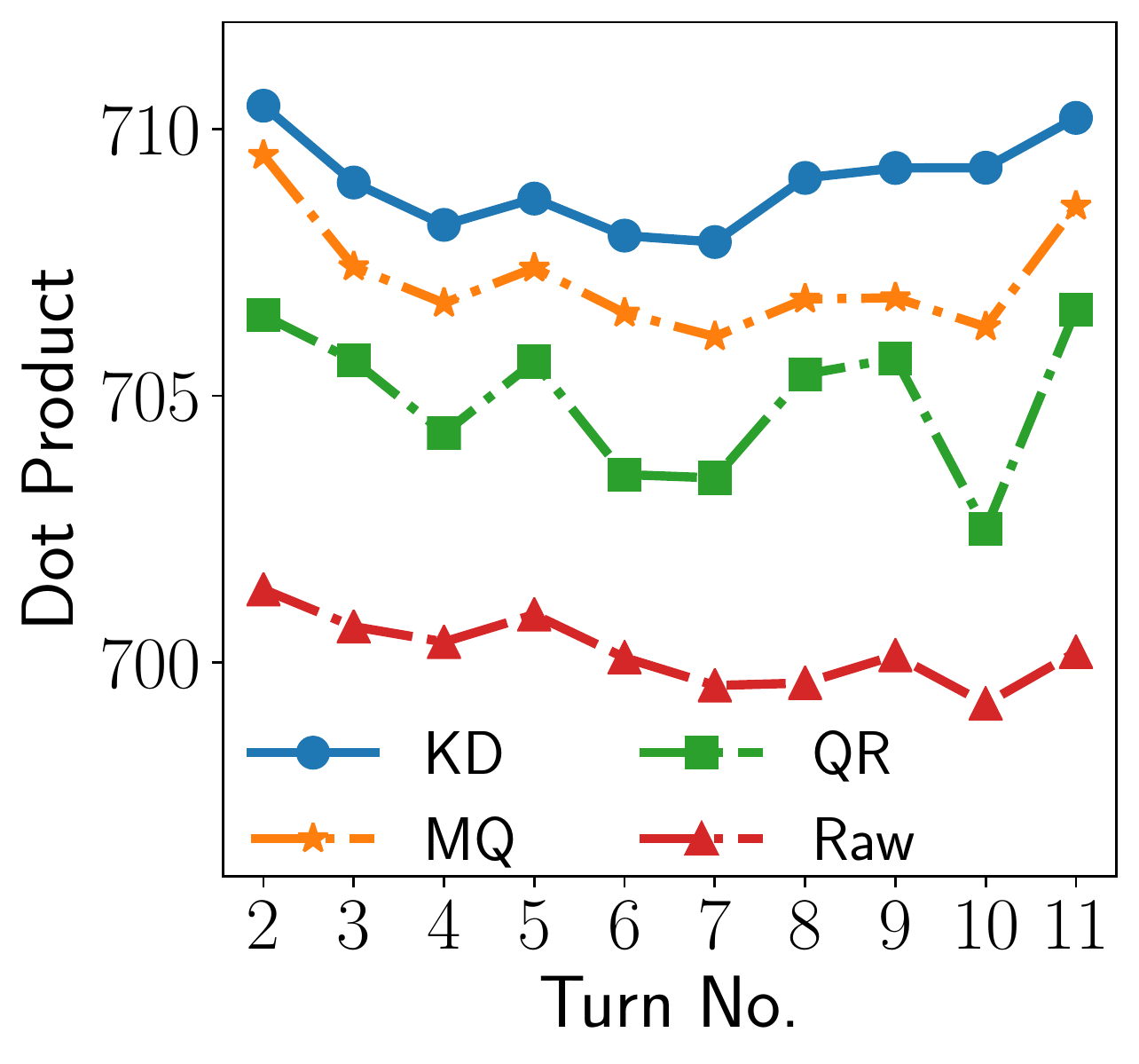}
        \caption{Similarity with previous turn. \label{fig:turn-sim}}
    \end{subfigure}
    \caption{Model behavior at different conversation turns on CAsT-19, including: \texttt{ANCE-Raw} \texttt{(Raw)}, \texttt{ANCE-Query} \texttt{Rewriter} \texttt{(QR)}, \texttt{\convance{} (KD)} and \texttt{ANCE-Manual} \texttt{(MQ)}. We evaluate their ranking accuracy at different turns (x-axes) in (a) and the similarity of query embeddings with previous round in (b).
 \label{fig:round}}
\end{figure}

%% file: Figures/fig-overlap.tex
\begin{figure*}[t]
    \centering
    \begin{subfigure}[t]{0.49\columnwidth}
        \centering
        \includegraphics[width=\linewidth]{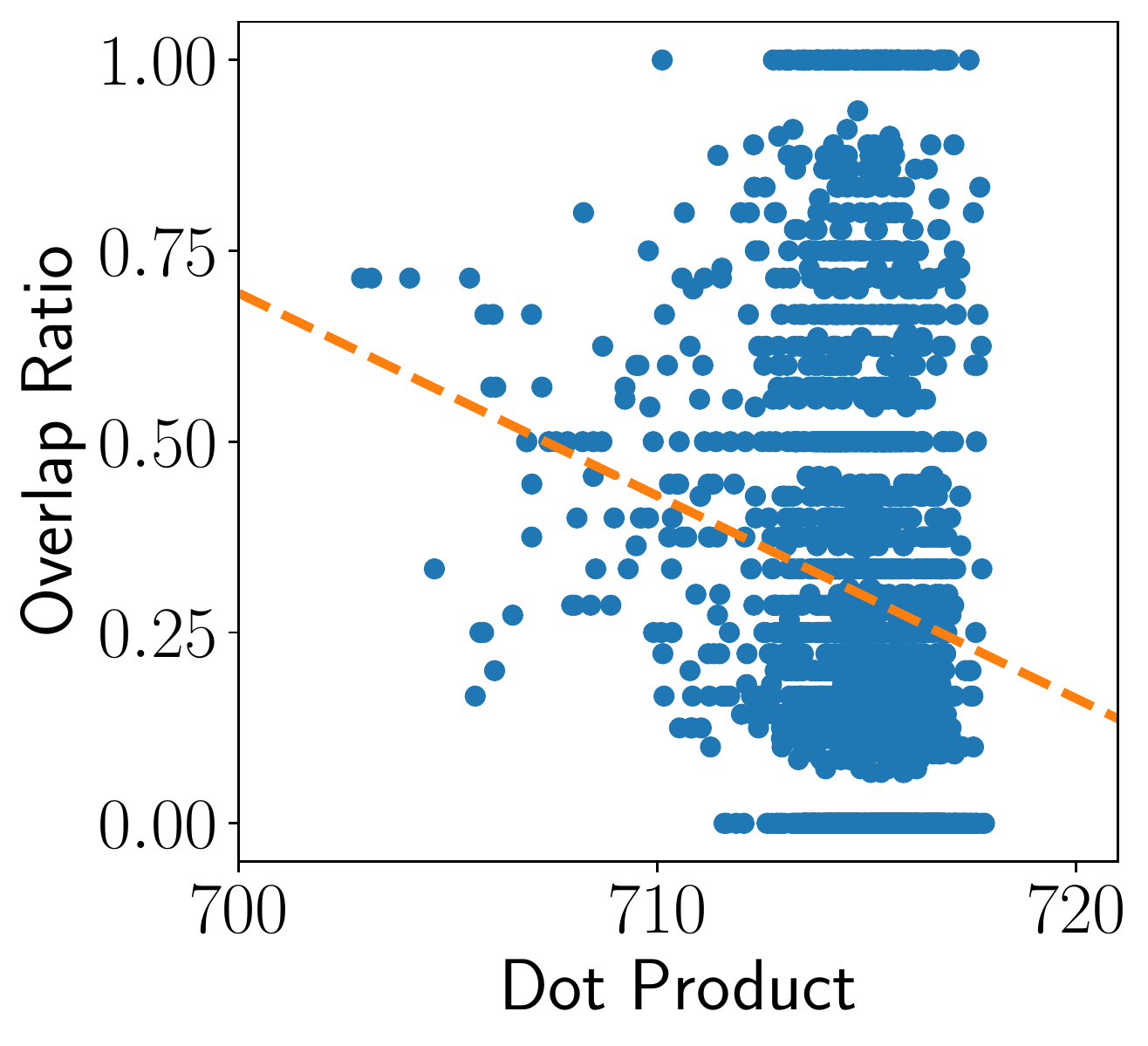}
        \caption{\texttt{\convance{} (Zero-Shot)}. \label{fig:overlap-zs}}
    \end{subfigure}
    \begin{subfigure}[t]{0.49\columnwidth}
        \centering
        \includegraphics[width=\linewidth]{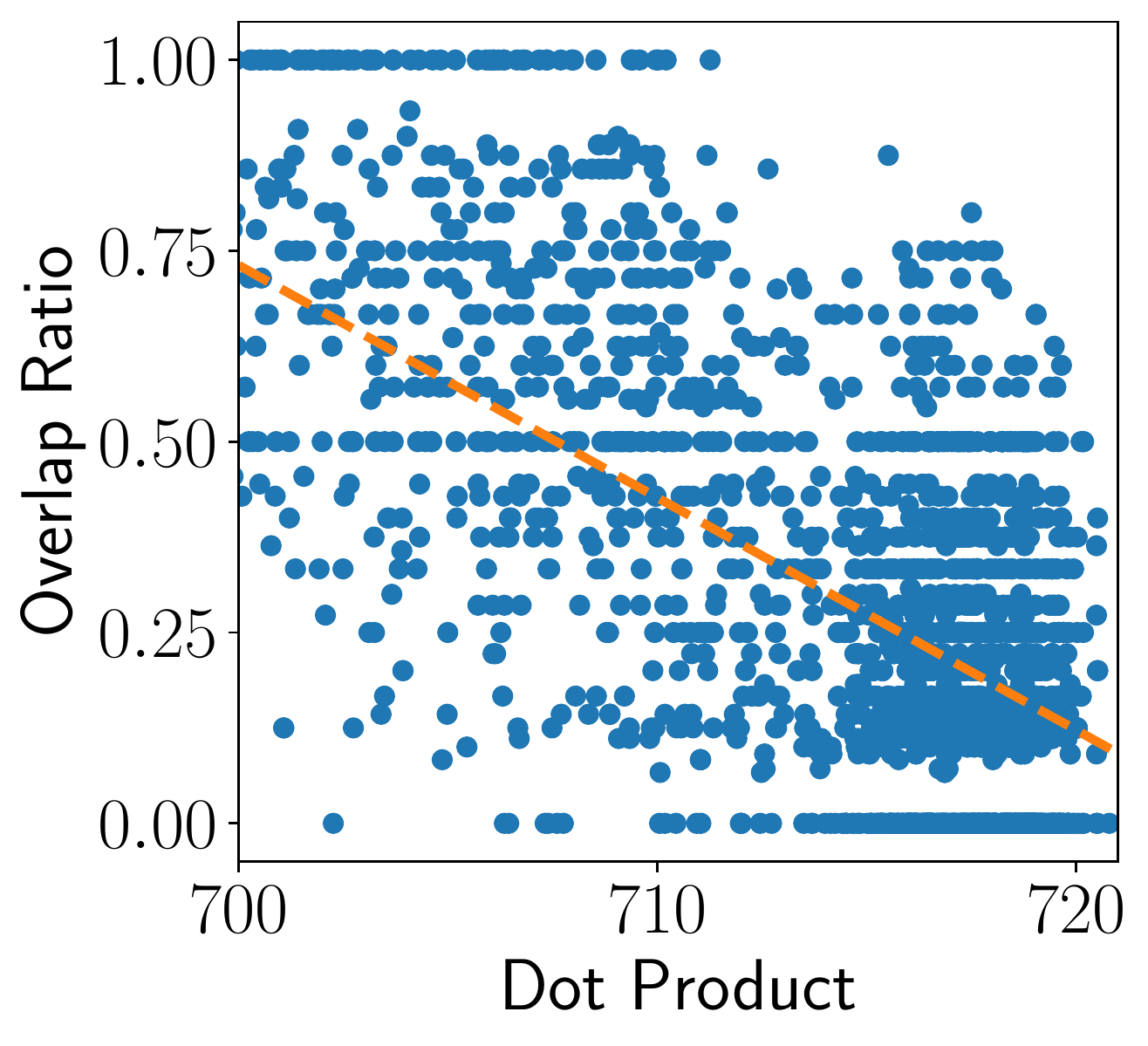}
        \caption{\texttt{ANCE-Query Rewriter}. \label{fig:overlap-qr}}
    \end{subfigure}
    \begin{subfigure}[t]{0.49\columnwidth}
        \centering
        \includegraphics[width=\linewidth]{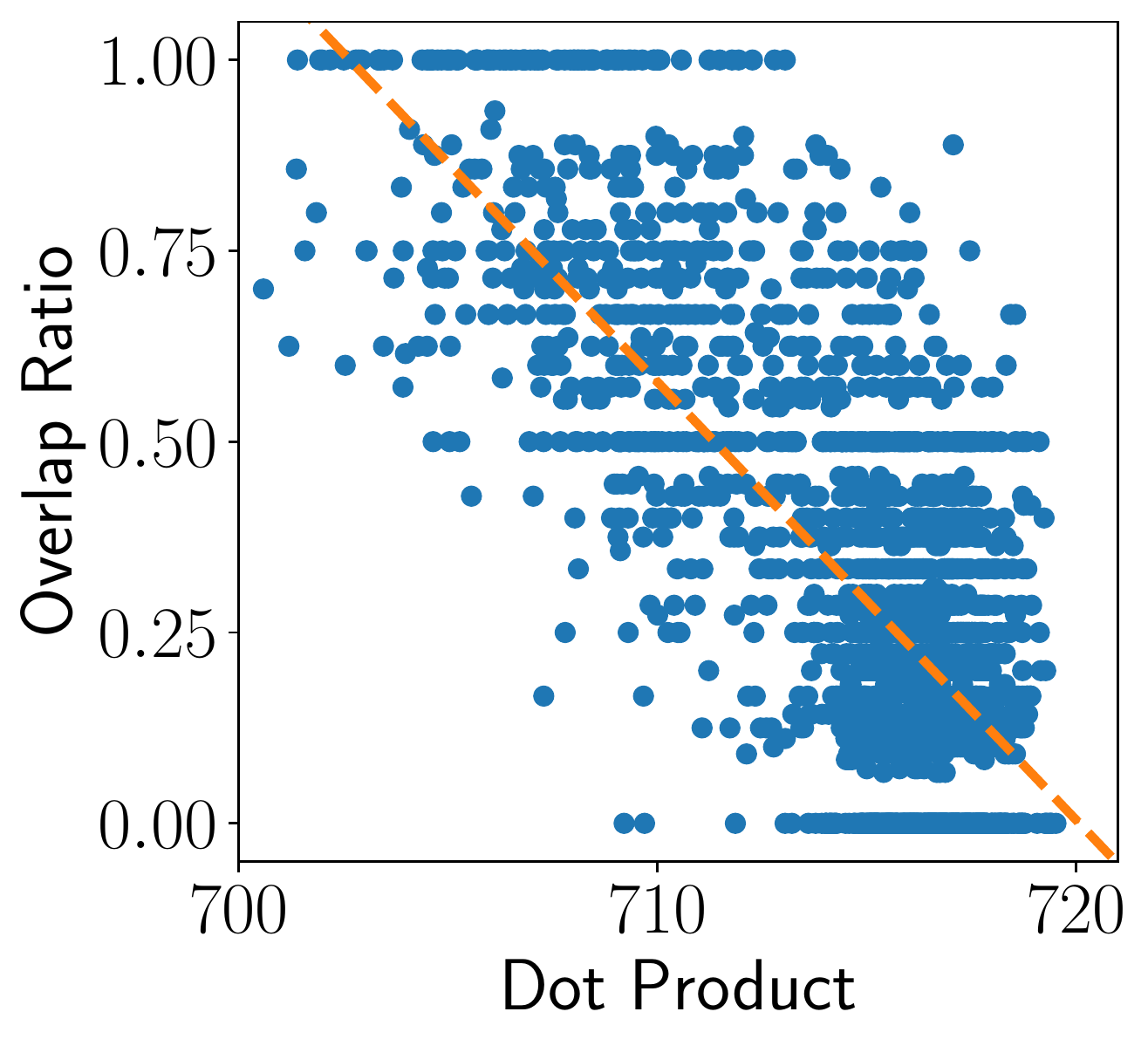}
        \caption{\texttt{\convance{} (KD)}. \label{fig:overlap-kd}}
    \end{subfigure}
    \begin{subfigure}[t]{0.49\columnwidth}
        \centering
        \includegraphics[width=\linewidth]{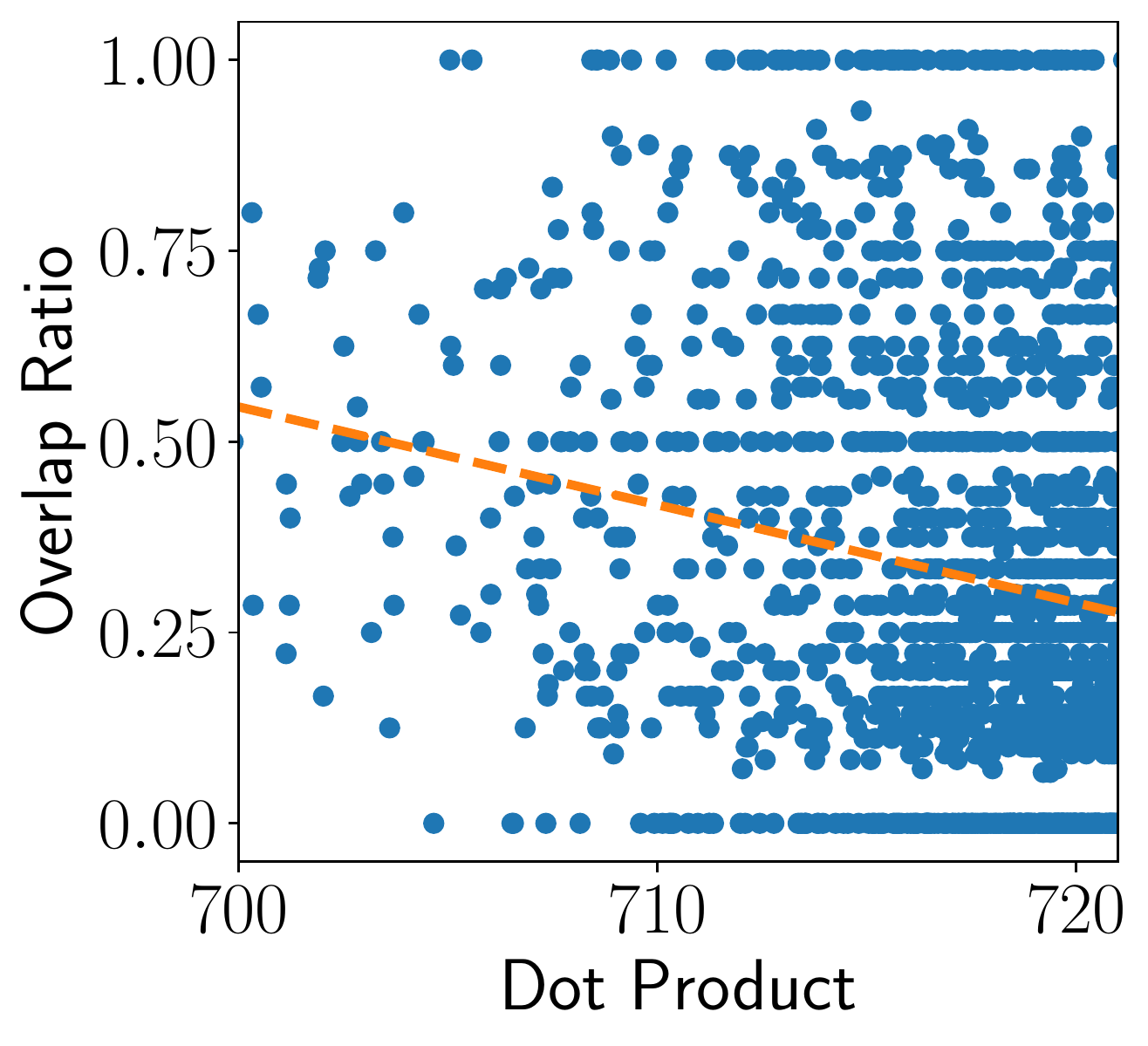}
        \caption{\texttt{\convance{} (Multi-Task)}. \label{fig:overlap-multi}}
    \end{subfigure}
    \caption{The overlap between query representations before and after discarding a previous conversation turn in the model input.
    The X-axes denote the similarity (dot product) score between the model's query embedding with and without that context turn. The Y-axes mark the token overlap ratio of the discarded previous turn with the manual oracle query of the current turn. All results are conducted on CAsT-19.
    \label{fig:overlap}}
    \Description{}
\end{figure*}

%% file: Tables/tab-case-new.tex
\begin{table*}
    \caption{Winning examples of \convance{} on CAsT-19. We show the historical queries, the current query (\underline{underlined}), its manual and automatic rewrites, and the first passages different methods disagree in the ranking results. The [bracketed] numbers are the score drops in \convance{} (KD) when the query is discarded. We manually find notable context terms and mark them \textit{italic}.
    }
    \label{tab:case} 
    \resizebox{0.99\linewidth}{!}{
    \begin{tabular}{p{0.48\linewidth}@{} |p{0.51\linewidth}} 
    \hline
    \textbf{Queries}         & \textbf{First Disagreed Passages}      \\
    \hline
    \multicolumn{2}{l}{\textbf{CAsT Topic-59}} \\ \hline
    Which weekend \textit{sports} have the most \textit{injuries?} {[-1.07]}  & \multirow{4}{*}{\begin{tabular}{@{}l@{}} \textbf{\convance{}}: An \textit{ACL} \textit{injury} is the tearing of the anterior cruciate \\ ligament ... \textit{ACL injuries} most commonly occur during \textit{sports} ... \\
    \textbf{\texttt{ANCE-Manual \& Query Rewriter}}: \textit{ACL}. Short for Access Control List,\\ an  \textit{ACL} is a list containing one or more ACE which are used by ...  \\ 
    \end{tabular}} \\ 
    What are the most common types of \textit{injuries}? {[-0.53]} &\\
    \underline{What is the \textit{ACL}?}  &\\
    \textbf{Manual \& Auto Rewrite}: What is the ACL?  & \\

    \hline
    \multicolumn{2}{l}{\textbf{CAsT Topic-37}}  \\
    \hline
 
    What was the Stanford Experiment? {[-1.51]} & \multirow{9}{*}{\begin{tabular}{@{}l@{}}
    \textbf{\convance{}}: In 1961, ... the Yale psychologist Stanley \textit{Milgram} performed \\... also known as the \textit{Milgram Experiment}, in order to determine ... \\ The \textit{Milgram Experiment} raised queries about the \textit{ethics} of ... \\
    \textbf{\texttt{ANCE-Query Rewriter}}: Forty years ago today, the Stanford Prison \\Experiment ... which gleaned powerful and unsettling insights into ... \\
    \textbf{\texttt{ANCE-Manual}}: The \textit{Milgram experiment} was a study ... that helped \\ measure a person's moral character. ... The \textit{Milgram experiment} \\ caused a huge amount of criticism among individuals. ...many were \\ completely convinced of the wrongness of what they were doing. \\
    \end{tabular}}\\
    What did it show? {[-0.10]} &\\
    Tell me about the author of the experiment. {[-0.22]} &\\
    Was it \textit{ethical}? {[-0.19]} &\\
    What are other similar experiments? {[-0.16]} &\\
    What happened in the \textit{Milgram experiment}? {[-3.92]} &\\
    \underline{Why was it important?}  &\\
                          
    \textbf{Auto Rewrite}: Why was the Stanford Experiment important? & \\
    \textbf{Manual Rewrite}: Why was the Milgram experiment important? & \\

    \hline
    \multicolumn{2}{l}{\textbf{CAsT Topic-56}}  \\
    \hline
    
    What is \textit{Darwin’s} theory in a nutshell? {[-0.77]} & \multirow{10}{*}{
    \begin{tabular}{@{}l@{}}
    \textbf{\convance{}}: \textit{Speciation} is the evolutionary process by which \\ \textit{reproductively} isolated \textit{biological} populations evolve to become distinct \\ species. ... \textit{Charles Darwin} was the first to describe the role of natural \\ selection in \textit{speciation}  in his 1859 book The Origin of Species. ... \\
    \textbf{\texttt{ANCE-Query Rewriter}}: ... \textit{Speciation} is defined by  ... as the distribution \\ of an element amongst defined chemical species in a system. Chemical \\ species include ... \\
    \textbf{\texttt{ANCE-Manual}}: \textit{Darwin's} Theory. In 1859, \textit{Charles Darwin} set out his \\ theory of evolution by natural selection as an explanation for adaptation \\and \textit{speciation}. ... \\
    \end{tabular}
    }\\
    How was it developed? {[-0.19]} & \\
    How do sexual and asexual \textit{reproduction} affect it? {[-0.10]} & \\
    How can fossils be used to understand it? {[-0.01]} & \\
    What is modern evidence for it? {[-0.12]} & \\
    What is the impact on modern \textit{biology}? {[-0.18]} & \\
    Compare and contrast microevolution and macroevolution. {[-0.37]} & \\
    \underline{What is the relationship to \textit{speciation}?}  & \\
    \textbf{Auto Rewrite}: What is the relationship to speciation? & \\
    \textbf{Manual Rewrite}: ... relationship of Darwin's theory to speciation? & \\
    
    \hline
    \end{tabular}
    }
\end{table*}

%% file: conclusion.tex
\section{Conclusion}

In this paper, we present a conversational dense retriever, \convance{}, that conducts the first stage retrieval of conversational search purely in the dense embedding space.
We propose a few-shot strategy that trains \convance{}'s query encoder to mimic the embeddings of manual oracle queries from a well-trained ad hoc dense retriever.
Our experiments on TREC CAsT and OR-QuAC demonstrate that \convance{} achieves state-of-the-art performance in first stage retrieval.
Detailed analysis shows that our simple approach can better capturing salient context information from the previous conversation turns in few-shot conversational search setting, as well as greatly improve online latency. 


\section{ACKNOWLEDGMENTS}
Part of this work is supported by the National Key Research and Development Program of China (No. 2020AAA0106501) and Beijing Academy of Artificial Intelligence (BAAI). 

\newpage